\documentclass[twocolumn]{aastex62}
\usepackage{xcolor}
\usepackage[utf8]{inputenc}

\newcommand{\thrirdfhl}{3FHL J1115$-$6117}
\newcommand{\fhl}{4FGL J1115.1$-$6118}
\newcommand{\fges}{FGES J1109.4$-$6115e}
\newcommand{\fgle}{4FGL J1109.4$-$6115e}
\graphicspath{{./}{figures/}}

\received{...}
\revised{...}
\accepted{...}
\submitjournal{ApJ}

\shorttitle{Gamma rays from NGC 3603}
\shortauthors{Saha et al.}

\begin{document}

\title{Morphological and spectral study of \fhl\ in the region of the young massive stellar cluster NGC 3603}
\email{labsaha@ucm.es, alberto.d@ucm.es}

\author[0000-0002-3171-5039]{L. Saha}
\affil{IPARCOS and Department of EMFTEL, Universidad Complutense de Madrid, E-28040 Madrid, Spain
\\
}%
\author[0000-0002-3433-4610]{\sc A. Dom\'inguez}
\affil{IPARCOS and Department of EMFTEL, Universidad Complutense de Madrid, E-28040 Madrid, Spain
\\
}%
\author[0000-0001-7523-570X]{L. Tibaldo}
\affil{IRAP, Universite de Toulouse, CNRS, UPS, CNES, Toulouse, France\\
}
\author[0000-0001-5544-0749]{S. Marchesi}
\affil{INAF - Osservatorio di Astrofisica e Scienza dello Spazio di Bologna, Via Piero Gobetti, 93/3, 40129, Bologna, Italy\\}
\affil{Department of Physics and Astronomy, Clemson University, Kinard Lab of Physics, Clemson, SC 29634-0978, USA\\
}
\author[0000-0002-6584-1703]{M. Ajello}
\affil{Department of Physics and Astronomy, Clemson University, Kinard Lab of Physics, Clemson, SC 29634-0978, USA\\
}
\author[0000-0002-4462-3686]{M. Lemoine-Goumard}
\affil{Centre d' Etudes Nucl\'eaires de Bordeaux Gradignan, IN2P3/CNRS, Universit\'e de Bordeaux, BP120, F-33175 Gradignan Cedex, France}
\author[0000-0002-8791-7908]{M. L\'opez}
\affil{IPARCOS and Department of EMFTEL, Universidad Complutense de Madrid, E-28040 Madrid, Spain
\\
}%



\begin{abstract}
We report a detailed study of an unidentified gamma-ray source located in the region of the compact stellar cluster NGC 3603. This is a star-forming region (SFR) powered by a massive cluster of OB stars. A dedicated analysis of about 10 years of data from 10 GeV--1 TeV, provided by the Large Area Telescope (LAT) onboard the \textit{Fermi Gamma-ray Space Telescope}, yields the detection of a pointlike source at a significance of 9$\sigma$.
The source photon spectrum can be described by a power-law model with best fit spectral index of $2.35 \pm 0.03$. In addition, the analysis of a deep \textit{Chandra} image in the 0.5--7\,keV band reliably rules out an extragalactic origin for the gamma rays. We also conclude that the broadband spectral energy distribution of the point source can be explained well with both leptonic and hadronic models. No firm evidence of association with any other classes of known gamma-ray emitters is found, therefore we speculate that \fhl\ is a gamma-ray emitting SFR.

\end{abstract}

\section{Introduction} \label{sec:intro}
Star-forming regions (SFRs) are considered potential contributors to the acceleration of Galactic cosmic rays, and detection of gamma rays from such SFRs in our Galaxy can establish the presence of relativistic charged particles. Indeed, the detection of gamma rays from the Cygnus cocoon indicates the presence of freshly accelerated high-energy particles in the SFR, making it the first case of a firm detection of such cosmic-ray acceleration \citep{Ackermann2011Sci}. However, the limited number of such gamma-ray detections has until recently prevented strong conclusions about the prevalence of SFRs as cosmic-ray sources.

 Gamma-ray emission in both the GeV and TeV energy ranges is detected from another massive stellar cluster, Westerlund 1. However, the origin of these gamma rays and their possible association with counterparts at other wavelengths remains poorly understood \citep{Abramowski_2012A&A, Ohm_2013MNRAS}. A recent study of Westerlund 1 shows a resemblance of the source's characteristics  to those of the Cygnus cocoon, indicating  cosmic-ray (CR) acceleration by the compact stellar cluster \citep{Aharonian2019NatureCom}. 
 In a search for such sources, \citet{Katsuta_2017} studied the region around Galactic coordinates, $l = 25^\circ$.0, $b = 0^\circ$.0 (dubbed as G25.0$+$0.0 region) using \textit{Fermi}-Large Area Telescope (LAT) data and other lower-energy  observations. A detailed study of this region likewise shows many similarities of the gamma-ray emission between G25.$0$+0.0 and the Cygnus cocoon. This, in turn, provides a hint that perhaps the same mechanism is at work in accelerating cosmic particles in these two regions. Diffuse extended gamma-ray emission at MeV--GeV energies was also found recently from the direction of the young massive star cluster Westerlund 2 \citep{Yang_Rui_2018}, although its association with the stellar cluster is not conclusive.  Therefore, SFRs  have become a new class of candidates for  CR accelerators \citep{Aharonian2019NatureCom}.

NGC 3603 is an SFR that has been observed at various wavebands, including gamma rays.  NGC 3603 is a nebula situated in the Carina spiral arm of the Milky Way at a distance of about $7 \pm 1$ kpc from the solar system  \citep{Melnick_1989,Crowther_1998MNRAS,dePree1999AJ,Pandey_2000PASJ,Nurmberger_2002A&A, Beccari_2010ApJ}. It is a massive (M $>$ 2000 $M_\odot$) H\textsc{ii} region and  one of the most luminous optically visible ones in the Milky Way, being powered by a cluster of OB stars \citep{Goss_1969ApL}. The H-alpha luminosity of NGC 3603 is $L(H_\alpha) \sim 1.5 \times 10^{39} \ \rm{ergs} \ \mathrm{ s^{-1}}$ \citep{Kennicutt_1984ApJ}, and the total mass of molecular clouds is about $4.1 \times 10^5 \rm M_\sun$ \citep{Grabelsky_1988ApJ}. The average age of the cluster is between 2 and 3 Myr  and the star formation in and around the young cluster has been going on for about 10--20 Myr \citep{Melnick_1989,Beccari_2010ApJ}.

In this paper, we investigate the SFR NGC 3603 using about 10 years of \textit{Fermi}-LAT data. A source positionally coincident with NGC 3603 was first detected in the third \textit{Fermi}-LAT Hard Source Catalog \citep[\thrirdfhl;][]{3FHL_2017ApJS}. Earlier studies above 10 GeV from the same region using 7 years of \textit{Fermi}-LAT data suggested the presence of extended emission \citep{Rui-zhi2017}.  The emission was claimed to have a hard spectrum with a photon index of 2.3 $\pm$ 0.1 from 1 GeV to 250 GeV.  The emission region was best-fitted with a Gaussian centred at RA (J2000) = 167$^\circ$.8 $\pm$0$^\circ$.1, Dec (J2000) = -61$^\circ$.3$\pm$0$^\circ$.1 with width = 1$^\circ$.1 $\pm$0$^\circ$.1, corresponding to a significance of more than $10\sigma$.   Here, we take advantage of an extended \textit{Fermi}-LAT dataset with better event-level analysis and improved interstellar emission models \citep{4FGL_2019arXiv}, along with X-ray data, to study emission from the fourth \textit{Fermi}-LAT catalog source \fhl.

The paper is organized as follows: we discuss the gamma-ray and X-ray analyses and their results in Section \ref{analysis} and Section \ref{sec:x-ray}, respectively. The modeling and interpretation are described in Section \ref{sec:modelling}. The results are discussed in Section \ref{sec:discussion}, while Section \ref{sec:conclusion} provides a summary of our findings.

\section{Observations and Data Reduction}
\label{analysis}
\subsection{Gamma-ray data}\label{sec:Gamma-ray data}
The Large Area Telescope (LAT) on board the \textit{Fermi Gamma-ray Space Telescope}  detects gamma rays in the energy range from 30 MeV to $>$ 500 GeV with a large effective area and a wide field of view \citep{Atwood2009}. In our analysis, we select nearly ten years (i.e., from 2008 September 1 to  2017 May 5) of Pass 8 SOURCE class (P8R3) LAT events in the reconstructed energy range from 300 MeV to 1 TeV within a 15$^{\circ}$ region of interest (ROI) around \fhl\ (associated with \thrirdfhl). 
The Fermi Science Tools (FSTs) analysis package\footnote{\url{https://fermi.gsfc.nasa.gov/ssc/data/analysis/software/}} version \texttt{v11r5p3} and the \emph{P8R3$_{-}$SOURCE$_{-}\!\!$V2} instrument response functions (IRFs) are used for the analysis. We also use a python-based package \texttt{Fermipy} (version 0.17.4)\footnote{\url{https://fermipy.readthedocs.io/en/latest/}} to facilitate analysis of data with the FSTs. The tool \texttt{gtselect} is used to select photons of energies greater than 300 MeV with arrival direction within 90$^\circ$ from the local zenith to remove contamination from the Earth's emission. The PASS 8 source class allows for the use of different \texttt{event types} based on the event-by-event quality of reconstructed direction (Point Spread Function; PSF) and energy. We split the data into four event types with associated PSF0, PSF1, PSF2 and PSF3, respectively, to avoid diluting high-quality events (PSF3) with poorly localized ones (PSF0).
The Galactic diffuse emission is modeled by the standard \textit{Fermi}-LAT diffuse emission model \textit{(gll\_iem\_v07.fits)}. The appropriate isotropic templates are also used following the FSTs documentation. We perform a binned maximum likelihood method to estimate the best-fit model parameters using a 15$^\circ \times 15^\circ$ square region centered on \fhl\ with 250 $\times$ 250 equally spaced spatial bins and 10 logarithmically spaced in energy. To enable the correction for the energy dispersion in the analysis, the flag \texttt{edisp} of \texttt{Fermipy} is set to \texttt{True} for all sources except for the isotropic emission.

 We first start with a baseline sky model within the ROI that includes all the 4FGL point sources, and all the \textit{Fermi}-LAT extended Galactic sources\footnote{These sources are obtained by \textit{Fermi-LAT} collaboration through a complete search for extended sources located within 7$^\circ$ from the Galactic plane, using 6 years of Fermi-LAT data above 10 GeV.} \citep[FGES;][]{FGES_2017ApJ...843..139A}  listed in the 4FGL catalog\footnote{\url{https://fermi.gsfc.nasa.gov/ssc/data/access/lat/8yr_catalog/gll_psc_v19.fit}}. The unassociated\footnote{The gamma-ray sources with no association with known classes of sources or the sources are confused or contaminated by the diffuse background.} point source \fhl\ (\thrirdfhl) is our source of interest in the ROI and the unassociated extended  source \fges\ ( associated with \fgle) from the FGES source catalog  is included in the model. The  FGES source \fges\ with an extension of 1$^\circ$.27 overlaps with 6 point-like 4FGL sources including \fhl. The center of the extended source is approximately 0.5$^\circ$ away from the point source \fhl. 
 Initially, we use the method \texttt{optimize} of \texttt{Fermipy} to find best-fit values for the spectral parameters of all models associated with the sources included in the ROI.
 After the initial optimization, we remove all sources for which the values of the predicted number of counts in the model, \textit{Npred}, are less than 2 and we free spectral shapes and normalizations for  all the sources which lie within 3$^\circ$ from the center of the ROI.  The isotropic template model is fixed to its value obtained after the first optimization of the ROI, but the normalization and index of the Galactic interstellar model is kept free for all different  models discussed below. Subsequently we fit the position of the source of interest. As the next step, we use an iterative maximum likelihood-based source finding algorithm to identify new point sources within 0.5$^\circ$ from the center of the ROI. The significance of each source is evaluated using a likelihood ratio test defined as TS = $\rm 2 log(\mathcal{L}_1/\mathcal{L}_0)$, where $\mathcal{L}_0$ and $\mathcal{L}_1$ are the likelihoods of the background model without the source (null hypothesis) and the hypothesis being tested (source plus background), respectively. The algorithm finds point sources within the ROI with TS $>$ 9. We continue searching for new sources until we do not find any source with TS $>$ 9. Following this, we remove all the sources with TS $ < $ 9 from the ROI and refit the values of all model parameters. All point sources with TS $>$ 9 obtained in these iterations are added to the baseline model, which we now call \textit{model A}. The energy range is chosen according to the particular study, morphological or spectral, as detailed below.

\begin{figure*}
\begin{tabular}{ccc}
\centering
\includegraphics[width=0.33\textwidth]{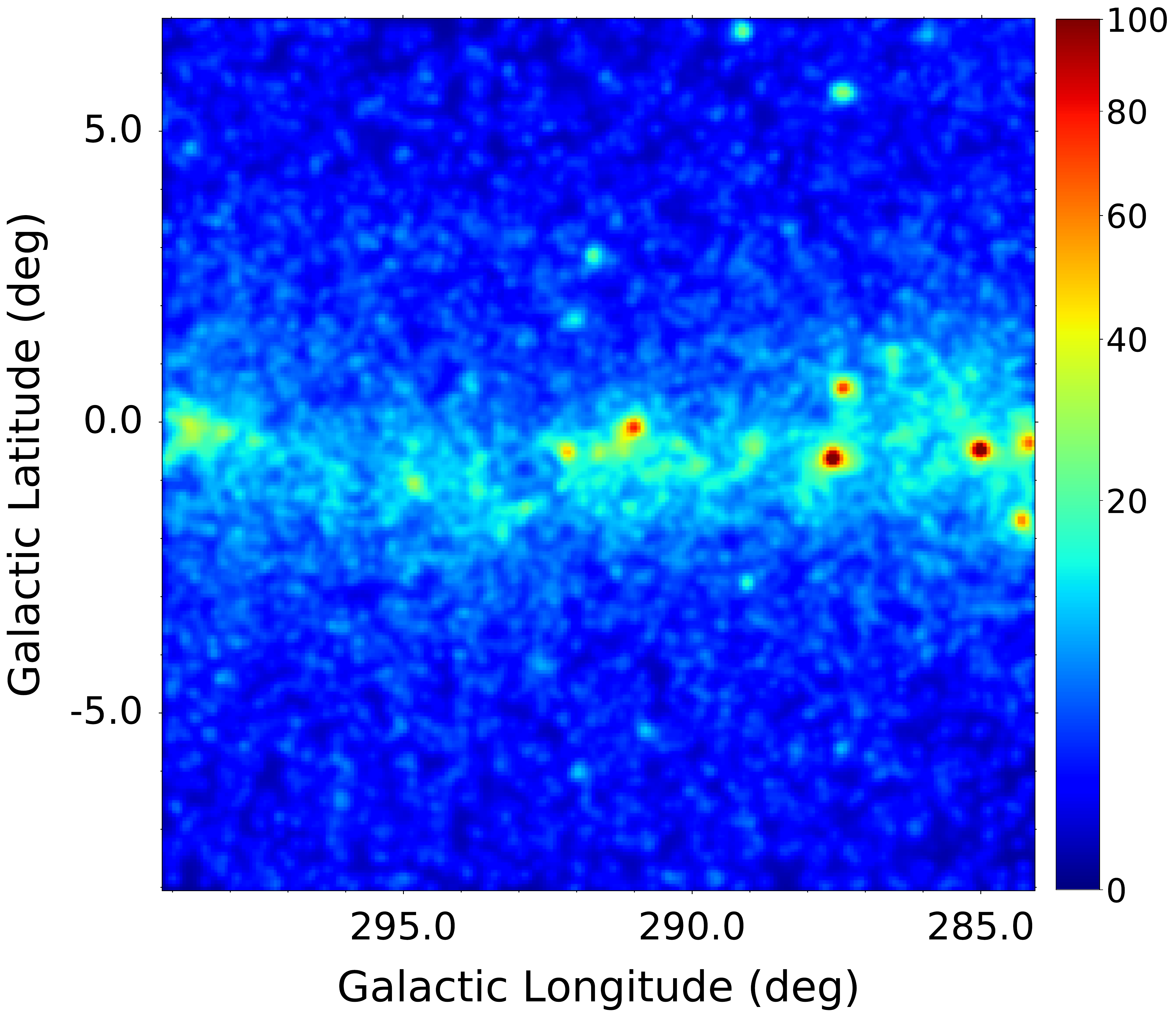}
\includegraphics[width=0.33\textwidth]{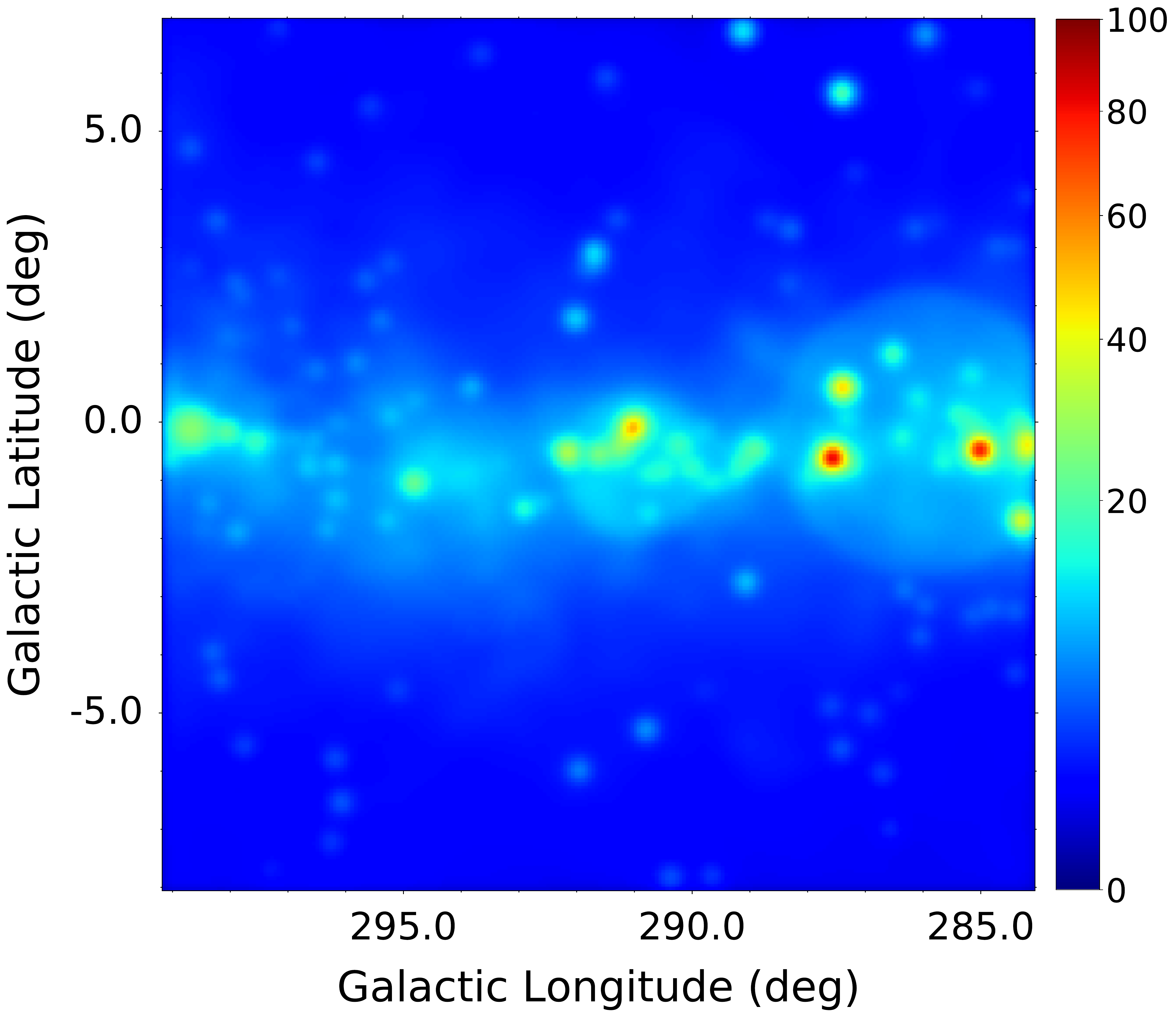}
\includegraphics[width=0.33\textwidth]{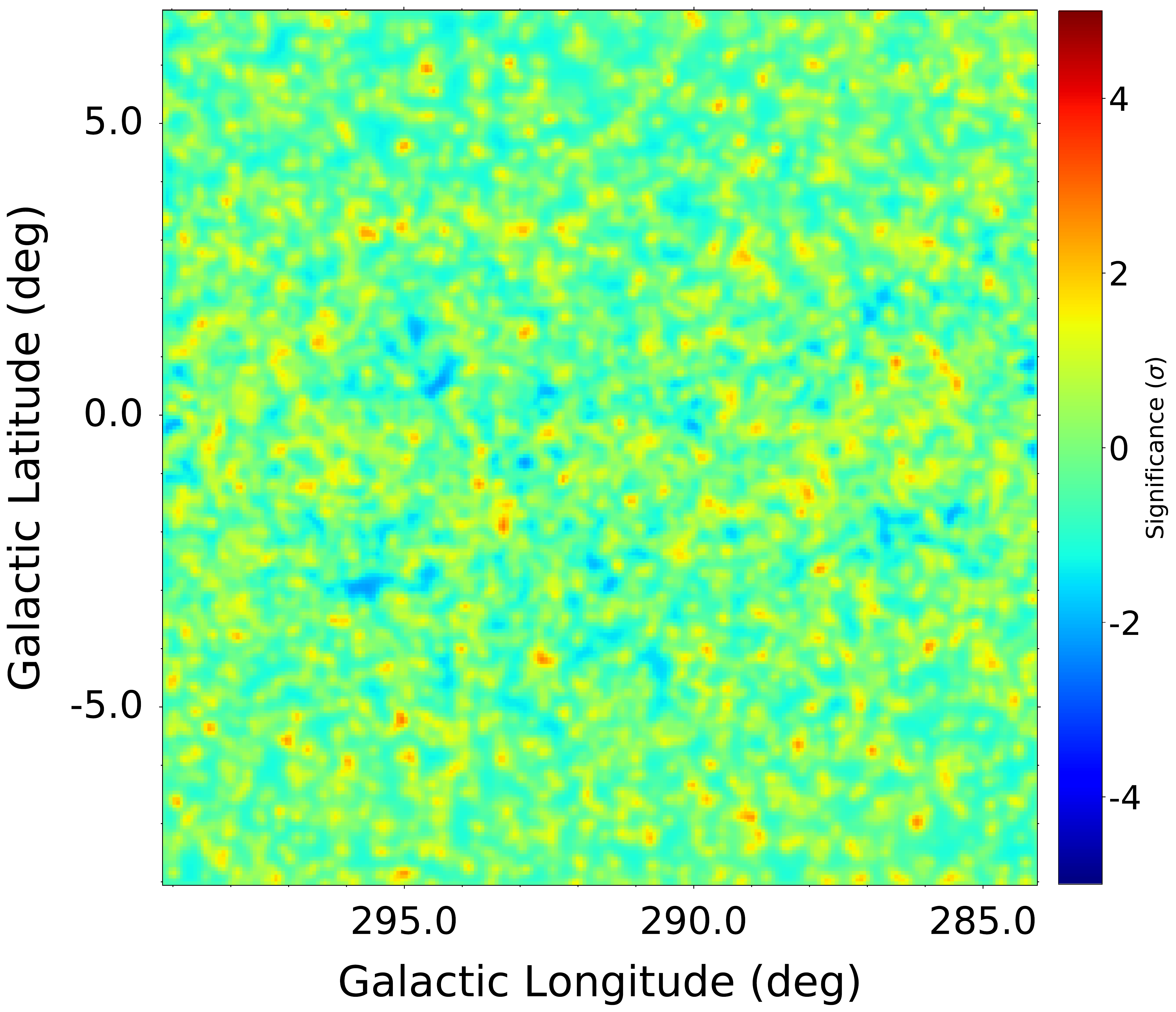}
\end{tabular}
\caption{Sky maps for \textit{model A}. Left) Data counts; middle) model counts; right) residual map. The source of interest i.e. \fhl\ is located at the center of the ROI. The field of view (FOV) is 15$^\circ \times 15^\circ$.\label{fig:skymaps_modelA}}
\end{figure*}

\subsubsection{Morphology}\label{sec:morphology}
 
For the morphological analysis of the source, only photons with energy 10 GeV--1 TeV are considered. The PSF of the \textit{Fermi}-LAT at high energies (above 10 GeV) is a factor of 10--15 better than that at 300 MeV. Moreover, the emission from pulsars is typically concentrated below 10 GeV, so using this energy range for morphological studies reduces the contamination from nearby pulsars within the ROI. 
The extension of a source is determined by calculating the TS of the extension, $\rm TS_{\rm ext} = 2log(\rm \mathcal{L}_{\rm ext}/\mathcal{L}_{pnt}$), where $\rm \mathcal{L}_{ext}$ is the maximum likelihood for an extended source model and $\rm \mathcal{L}_{pnt}$ for a pointlike source model.
To find extension of the source, we adopt radially symmetric Gaussian model for characterizing it, varying its $\sigma$ parameter from 0.01$^\circ$ to  1.5$^\circ$ in steps of 0.1 degrees. We also simultaneously leave the centroid of the source free within the 1$\sigma$ extension of the Gaussian. First, we re-optimize the model for \fges, finding it best localized at RA = 167$^\circ.35 \pm   0^\circ.10$ and DEC =$-$61$^\circ.25 \pm 0^\circ.07$  with an extension of $0.95^\circ \pm 0^\circ.06 $ with TS$_{\rm ext}$ = 146.4. For comparison, in the FGES catalog the extension for a 2D-Gaussian model is found to be 0$^\circ$.88 $\pm 0^\circ.05$  with RA = 167$^\circ$.36 $\pm$   0$^\circ$.05 and  DEC=-61$^\circ$.26 $\pm$   0$^\circ$.04. We update the model with this extension before fitting for \fhl. Next, we find that the  maximum likelihood value is obtained when \fhl\ is centered at RA = 168$^\circ$.78 $\pm$   0$^\circ$.01 and  DEC=-61$^\circ$.29 $\pm$   0$^\circ$.02 with an extension of 0$^\circ$.08 $\pm$ 0$^\circ$.02 with TS$_{\rm ext}$ = 7.7, which is well below the value of about 16 ($\sim$4$\sigma$) for claiming this as an extended source. According to the 4FGL catalog, \fhl\ is located at RA = 168$^\circ$.77 $\pm$ 0$^\circ$.02 and  DEC=-61$^\circ$.30 $\pm$   0$^\circ$.02. Hence, the results obtained here for \fhl\ are consistent with the 4FGL catalog values.

Since the stellar cluster lies in a complex region with many nearby/overlapping sources in the 4FGL catalog, we test how the results depend on the sources assumed for the ROI. In \textit{model B}, we  remove the extended \fges\ from the \textit{model A} and refit the model to the data and estimate the maximum log-likelihood value.  In \textit{model C}, we instead remove unassociated sources within 3$^\circ$ from the center of the ROI but  keep \fges\ and \fhl. This allows us to understand the impacts of other unassociated sources compared in modeling this region. Finally, in \textit{model D}, we remove all the unassociated sources including \fges\ from the 3$^\circ$ region of the center of the ROI.

\begin{figure}
\includegraphics[width=0.45\textwidth]{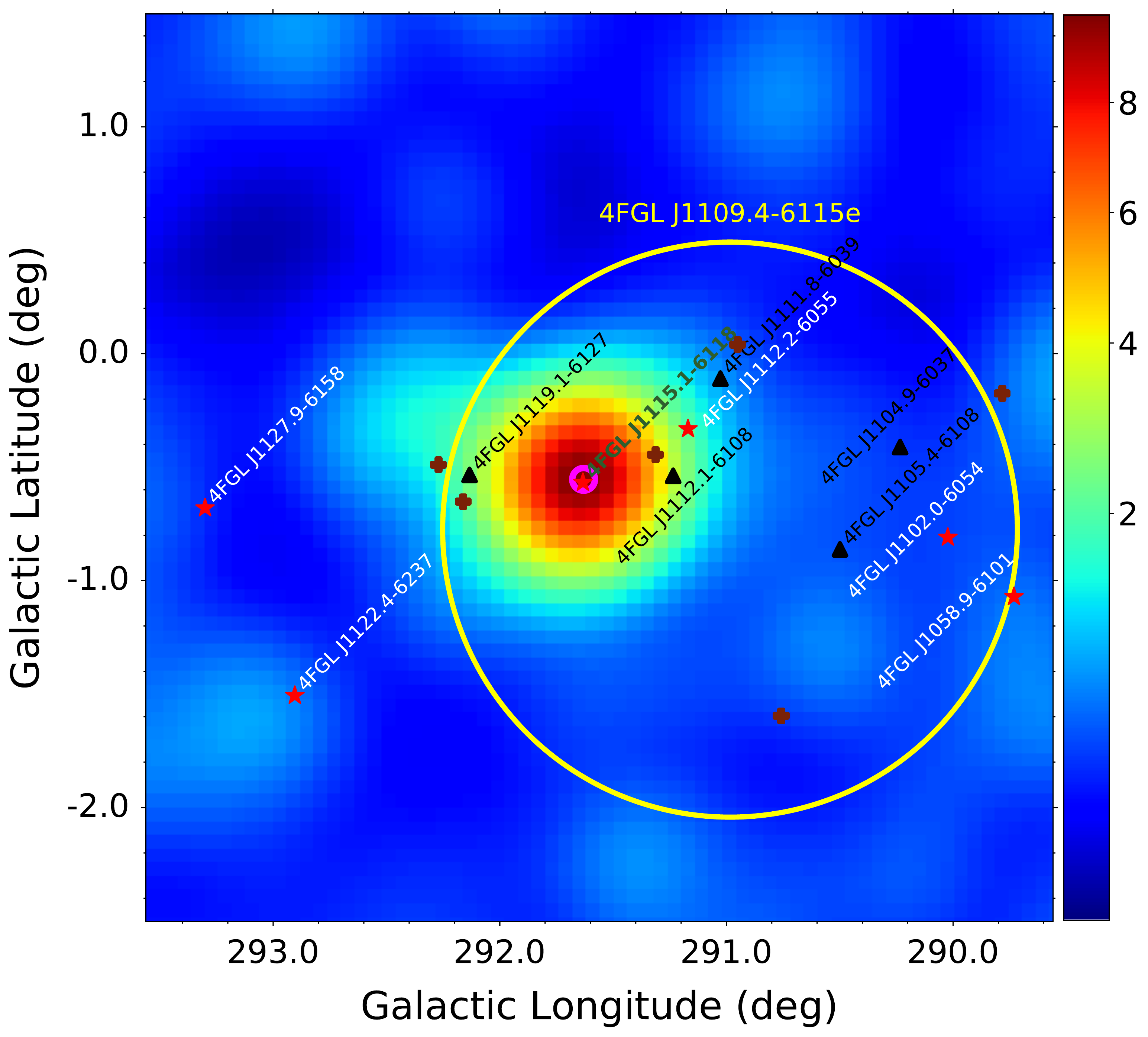}
\caption{The TS map of \fhl\ for energy 10 GeV - 1 TeV. The map is obtained for \textit{model A} excluding the source of interest from the model. The associated 4FGL catalog point sources present in this region are shown with black `triangle' marks. The unassociated 4FGL point sources are marked with Red star marks, whereas the unassociated extended \fges\ is shown with a yellow circle. The 'star' within the magenta circle indicates the location of \fhl\ according to the 4FGL catalog, whereas the magenta circle indicates the 95\% positional uncertainty radius of the point source \fhl\  obtained from the results of the analysis presented here. Other point sources obtained through point source seraching procedure are shown with filled brown 'plus' markers. \label{fig:skymaps_sources}}
\end{figure}

The log-likelihood values for these different cases are given in Table \ref{tab:likelihood_values}. We also estimate the Akaike criterion (AIC) for each  model, defined as $$AIC = 2k - 2 log \mathcal{L},$$ where k is the number of free parameters in the model and $\mathcal{L}$ is the likelihood value of the model. It is evident from Table \ref{tab:likelihood_values} that the model with the point source \fhl\, i.e. \textit{model A}, provides the maximum log-likelihood value and the minimum AIC, which makes \textit{model A} the  preferred one.  Fig. \ref{fig:skymaps_modelA} shows the counts map for data and model and the residual map within a region of 15$^\circ \times 15^\circ$ for \textit{model A}, which indicates that the region is satisfactorily modelled. We also check the residual maps for the other models (\textit{B}, \textit{C}, and \textit{D}) and find unmodelled emission, further supporting the conclusion that \textit{model A} is the preferred one. Therefore, this establishes the fact that \fhl\ is not significantly extended. We additionally show the TS map for \textit{model A} in a region of $4^\circ.0 \times 4^\circ.0$ along with the sources present in this region in Fig. \ref{fig:skymaps_sources}.

As mentioned earlier, the gamma-ray source toward NGC 3603 was found  to be significantly extended above 10 GeV in the study by \citet{Rui-zhi2017}. However, our results favor \textit{model A}, in which the source of interest is not significantly extended. By comparing \textit{model A} with models \textit{B} and \textit{D}, we see that the extension of the source coincident with the stellar cluster critically depends on the details of the modelling of the surrounding region, and in particular on the description of \fges. This extended source \fges\ was not present in the analysis by \citet{Rui-zhi2017}.
Furthermore, we investigate the reason for this mismatch by repeating the same analysis procedure and using the same set of data and background models\footnote{Here all the models are from \textit{Fermi}-LAT 3FGL catalog \url{https://fermi.gsfc.nasa.gov/ssc/data/access/lat/4yr_catalog/gll_psc_v16.fit}. 
} as mentioned in \citet{Rui-zhi2017}. We find that the data favor a significant source extension. However, when we use the new version of the Galactic interstellar emission model, i.e., \textit{gll\_iem\_v07.fits} as used for our analysis presented here, we find that the source appears extended but less significantly ($\rm TS_{ext} = 17$) than that obtained with the old version ($\rm TS_{ext} = 57$) of the Galactic diffuse model (\textit{gll\_iem\_v06.fits}). When we combine the new Galactic model along with \fges, we get results similar to those for \textit{model A}, i.e., the source appears to not be significantly extended. This indicates that the Galactic interstellar emission model and inclusion of the extended source plays a crucial role in the determination of the properties of the gamma-ray source toward NGC 3603. We also find that the new Galactic diffuse model provides a better global fit in this region than the old model for this particular part of the sky. The difference of log-likelihood values for these two cases is $\Delta \log \mathcal{L} = 101$.

\subsubsection{Spectrum}\label{sec:spectrum}
For the spectral study, we consider data for the energy range from 300 MeV--1 TeV. We characterize the spectral energy distribution (SED) of \fhl\ using the best model, i.e., \textit{model A} as discussed in section \ref{sec:morphology}.
We performed a spectral fit over the entire energy range using, first, a power-law spectral shape defined as follows:

$$\rm PL: {dN \over dE} = N_0 \left({E\over E_0}\right)^{-\alpha}.$$

\begin{figure}
\includegraphics[width=0.49\textwidth]{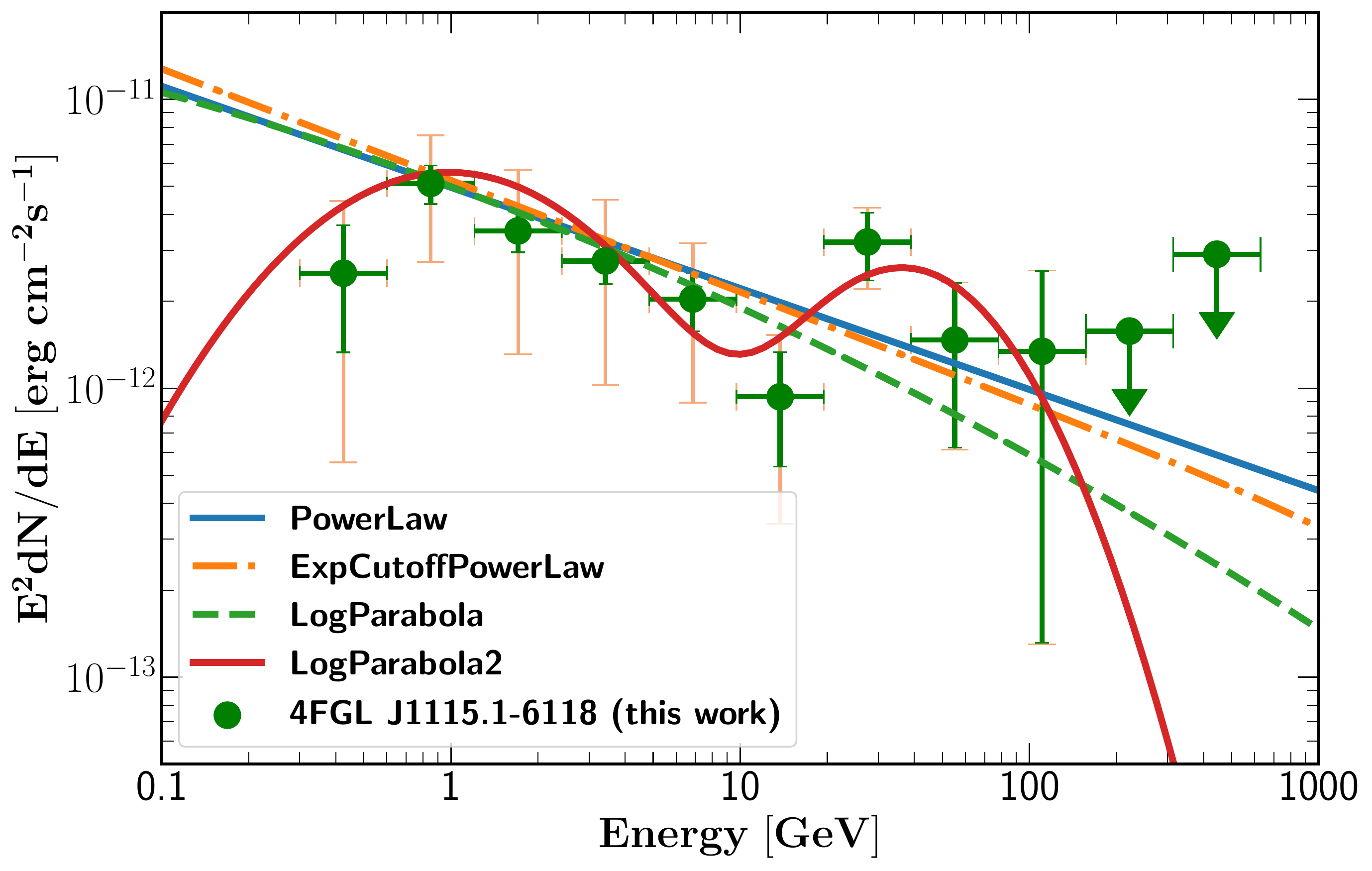}
\caption{Gamma-ray spectral energy density (SED) of \fhl\ measured by \textit{Fermi}-LAT in the energy range of 300 MeV--1 TeV. The SED is fitted with different spectral shapes. The results of the fit for different models are shown in Table \ref{tab:models}. The data points correspond to the power-law model. The LP2 model (solid red line) corresponds to sum of two LogParabola models, with peaks fixed at 700 MeV and 20 GeV, respectively. The total errors (systematic and statistical) are shown with orange lines. \label{fig:sed_fit_test}}
\end{figure}

We then estimated the SED (shown in Fig. \ref{fig:sed_fit_test}) by varying the normalization of this model independently in 11 energy bins spaced uniformly in log space from 300 MeV to 1 TeV.

The SED suggests that the emission has two different components, one below 10 GeV and the other above 10 GeV. In order to understand the significance of the spectral curvature we fit the data with different spectral shapes, including a ``custom'' model which can account for the two-peaked nature of the spectrum. This special function is essentially a sum of two LogParabolas (hereafter LP2) models with peaks fixed at 700 MeV and 20 GeV, respectively. In addition, we consider a LogParabola (LP) and a power-law with exponential cutoff (ECPL) model. The models are defined as follows.

$$\rm LP: {dN \over dE} = N_0 \left({E\over E_0}\right)^{-(\alpha + \beta \log(E/E_0))} $$
$$\rm ECPL: {dN \over dE} = N_0 \left({E\over E_0}\right)^{-\alpha} \exp\left(-{E \over E_{cut}}\right),$$

$$\rm LP2: {{dN \over dE}} = \rm N_0 \left({E \over E_1}\right)^{-(\alpha_1 + \beta_1log(E/E_1))}$$

$$~~~+\rm N_1 \left({E\over E_{2}}\right)^{-(\alpha_2 + \beta_2log(E/E_{2}))},$$

\noindent
where $\rm N_0$, $\alpha$, $\beta$, $\beta_1$, $\beta_2$, $\rm E_{cut}$, $\rm N_1$, $\alpha_1$, $\alpha_2$ are parameters of the models. The fitted models are shown in Fig. \ref{fig:sed_fit_test} and the results of the fits are shown in Table \ref{tab:models}. It is evident from the table that the PL model is better than its nested counterparts, ECPL and LP. The largest improvement w.r.t. to the PL model is given by LP2 model with TS = 14.2 for 4 additional degrees of freedom (d.o.f.), which corresponds to an improvement of only 2.7$\sigma$. We therefore conclude that the spectral curvature is not particularly significant.

Hence, for our present study, we adopt the power-law model for the rest of the discussion. The best-fit spectral photon index of PL model is, $\alpha$ = $2.35  \pm 0.03$ and the total integrated flux is found to be $F(>300 ~\rm MeV) = (1.2 \pm 0.1) \times 10^{-8} ~\rm photons ~\rm cm^{-2} ~\rm s^{-1}$.

In order to calculate the systematic errors associated with the Galactic diffuse models, we repeat the procedure of estimating the SED using an older diffuse model \textit{(gll\_iem\_v06.fits)}. Before estimating the SED, we fix the extension of both 4FGL J1115-6118 and \fges\ obtained with the new diffuse model \textit{(gll\_iem\_v07.fits)}. In addition to the uncertainty of the Galactic diffuse background model, we also consider the systematic uncertainty associated with the uncertainties of the effective area of the detector\footnote{\url{https://fermi.gsfc.nasa.gov/ssc/data/analysis/LAT_caveats.html}}. The total errors (systematic and statistical) on the SED are also shown in Fig. \ref{fig:sed_fit_test}.

\begin{figure}
\includegraphics[width=0.49\textwidth]{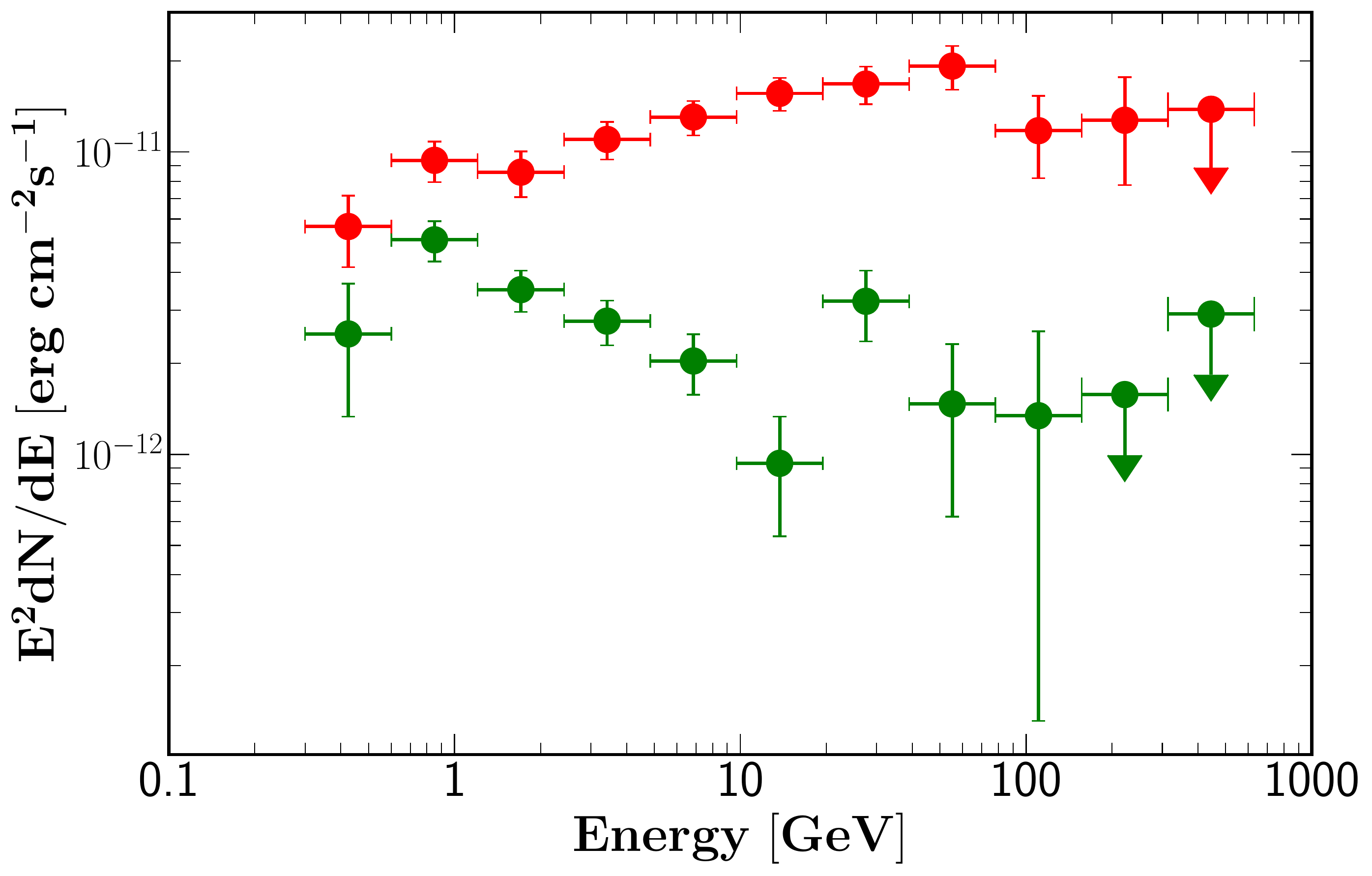}
\caption{ SEDs of the source of interest \fhl\ and the extended source \fges\  in the energy range of 300 MeV--1 TeV. The error bars shown are statistical only.}\label{fig:sed_point_extended}
\end{figure}

As mentioned in the previous section, the treatment of the extended diffuse emission region \fges\ plays a crucial role in the observed source characteristics. Hence, we also estimate its SED to examine the spectral properties of this extended source. The observed spectrum is shown in Fig. \ref{fig:sed_point_extended}. The two spectra are quite different. It appears that the nature of \fhl\ is different from \fges, and the point source is not contributing to the estimated flux of the extended source.  Note that \fges\ is flagged in the FGES catalog as confused and possibly contaminated by the diffuse background. Nevertheless, the relationship between the extended source \fges\ and the point-like source seen at the center of the stellar cluster, \fhl, deserves further investigation.

\renewcommand*{\thefootnote}{\fnsymbol{footnote}}
\begin{table*}[]
    \centering
     \caption{Results of maximum-likelihood fit for different models. The values of TS$_{ext}$, RA, DEC and extension are associated with the source of interest \fhl. The details of the models are given in Section \ref{sec:morphology}}.
    \begin{tabular}{c|c|c|c|c|c|c|c|c}
    \hline 
    \hline
        Models    & $\Delta \log\mathcal{L}$\footnote{Calculated w.r.t. \textit{model A}} & d.o.f             & TS$_{ext}$ & \multicolumn{2}{c}{Best-fit location}   &  Best-fit extension  & $\Delta$ AIC$^a$ & 95\% C.L. Upper Limit \\
                  &                  &                   &         & RA (deg)    & DEC(deg) & (deg)              &  &on extension (deg)\\     
    \hline    
       model A    &  0             & 37   & 7.7    & 168.78 $\pm$ 0.01 & -61.29 $\pm$ 0.02 & 0.081$^{+0.024}_{- 0.023}$   & 0 & 0.12\\
       model B    & -37            & 35   & 49.7   & 167.94$\pm$ 0.07 &  -61.30 $\pm$ 0.06 & 0.966$^{+0.069}_{- 0.067}$  & 70 & -\\
       model C    & -31            & 29   & 7.5    & 167.97 $\pm$ 0.01 & -60.68 $\pm$ 0.01 & 0.134$^{+0.017}_{- 0.016}$  & 46 & 0.16\\ 
       model D    & -76            & 27   & 74.1   & 168.23 $\pm$ 0.04&  -61.11 $\pm$ 0.03 & 0.903$^{+0.065}_{- 0.066}$  & 132 & -\\  
      \hline 
    \end{tabular}
    
    \label{tab:likelihood_values}
\end{table*}

\begin{table}
    \centering
     \caption{Significance of the spectral curvature at MeV--GeV energies for different spectral shapes.}
    \begin{tabular}{c|c|c|c}
    \hline
    \hline
        Spectral model      & $\Delta \log\mathcal{L}$\footnote{Calculated w.r.t.\textit{model A}} & d.o.f & $\Delta$ AIC$^a$ \\
    \hline    
       Powerlaw             &  0.0          &  22   &  0.0  \\
       ExpCutoffPowerlaw    &  0.6          &  23   & 0.8   \\
       LogParabola          &  1.0          &  23   & 0.0    \\
       LogParabola2         &  7.1          &  26   & -6.2  \\
      \hline 
    \end{tabular}
    \label{tab:models}
\end{table}
\renewcommand*{\thefootnote}{\arabic{footnote}}
\setcounter{footnote}{0}

\section{X-ray association} \label{sec:x-ray}
\noindent
To improve our understanding of \fhl\ we study the 0.5--7\,keV emission of NGC 3603 as observed with \textit{Chandra} (ObsID: 12329; the observation was taken on October 15, 2010, and is 150\,ks long). We reduce the \textit{Chandra} data using the CIAO \citep{fruscione06} 4.9 software and the \textit{Chandra} Calibration Data Base (\texttt{caldb}) 4.8.2, adopting standard procedures. The previous works used the excellent spatial resolution of \textit{Chandra} to identify several hundreds \citep{moffat02} and even $>$1000 point-like sources \citep{townsley11} with relatively short observations (46\,ks). \citet{townsley14} further studied this region with deepest exposure (490\,ks) using \textit{Chandra} data which showed even $>$ 4000 X-ray point sources. We, however, reanalyze this observation to look for non-thermal X-ray emission from bright X-ray sources. The  \texttt{wavdetect} tool of CIAO is used to determine how many \textit{Chandra} objects are located within the 95\% localization uncertainty radius of \fhl. Fig. \ref{fig:x-ray_map} shows the X-ray counts map. The details of the analysis are given in Appendix \ref{sec:x-ray_table}.

The X-ray analysis shows that there are 38 bright point sources within the region with net counts above 70. The spectral analysis shows that 31 out of 38 sources are best-fitted with a thermal \texttt{mekal} model,  which is commonly adopted to fit the X-ray spectra of young, bright stars \citep{mewe85,mewe86,liedahl95}. The other seven objects are best-fitted with a power-law model. It is also found that most of the X-ray bright sources are of Galactic origin due to the fitted parameter N$\rm_H$ being very close to that of the Galactic one (N$_{\rm H,Gal}$=1.16 $\times10^{22}\,\rm cm^{-2}$). The integrated X-ray flux in the 2--10 keV band for the whole region corresponding to \fhl\ is $(3.3 \pm 0.2) \times 10^{-12}~ \rm erg ~cm^{-2}~ s^{-1}$. The sources with non-thermal spectra are expected to be associated with the source of our interest.  However, given the relatively large size of the \fhl\ region with respect to the \textit{Chandra} spatial resolution, any strong correlation between X-ray and gamma-ray emission cannot be established.

\begin{figure}
\includegraphics[width=0.45\textwidth]{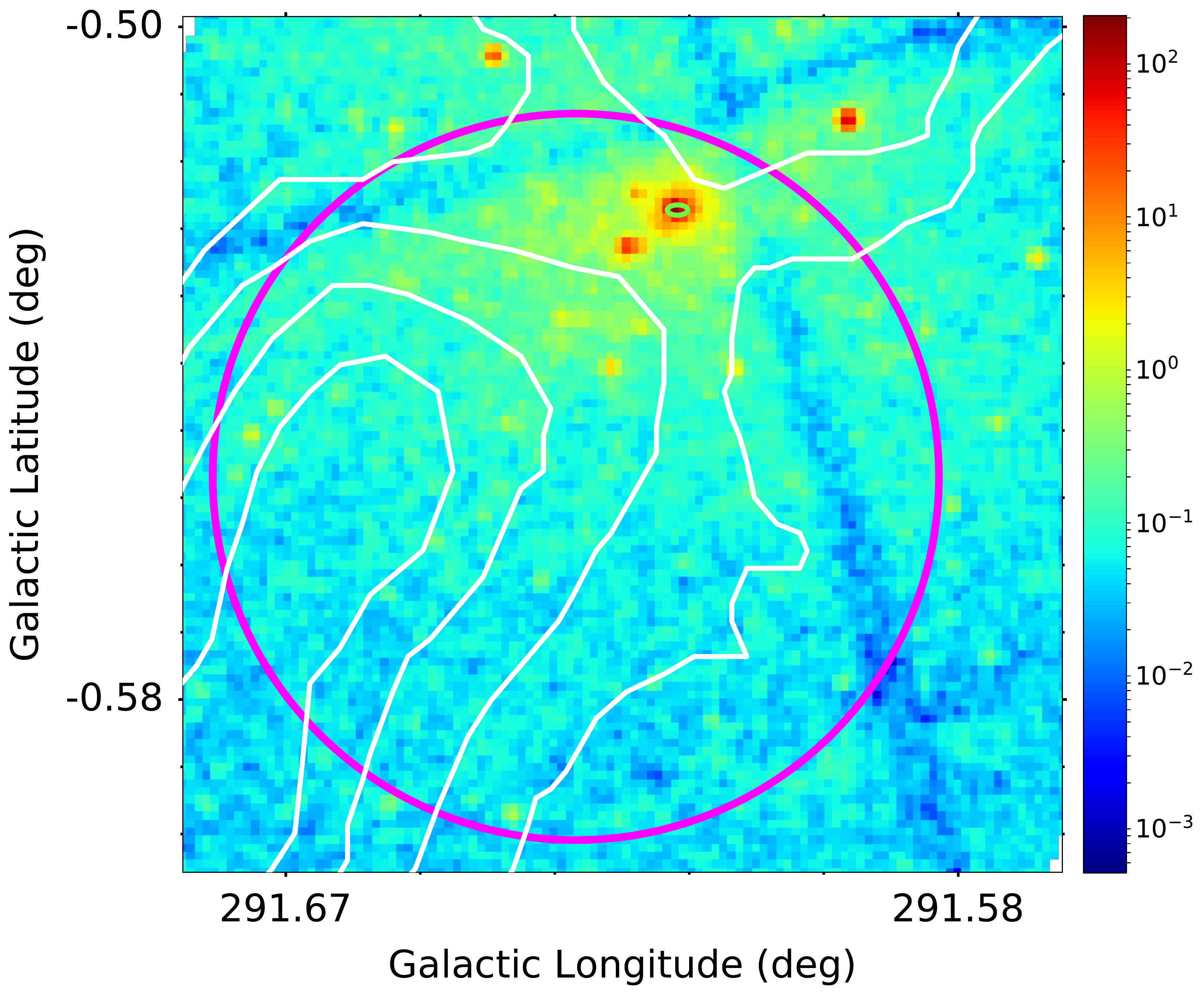}
\caption{X-ray counts map from the direction of NGC 3603. The circle in magenta shows a 95\% confidence localization uncertainty of the Fermi-LAT point source \fhl\ obtained from the \textit{Fermi}-LAT analysis. The region bounded by the magenta circle except the central part of the cluster marked with a green small ellipse is considered for the analysis of X-ray data. It is evident from the figure that the observed bright X-ray emission is well within the \textit{Fermi}-LAT error circle. The MC contours (white solid lines) are also overlaid.}\label{fig:x-ray_map}
\end{figure}

\section{Modelling the spectral energy distribution}\label{sec:modelling}
The observed gamma-ray fluxes at MeV--GeV energies may be interpreted based on leptonic and hadronic emission models. To find out the dominant emission mechanisms for \fhl, we consider inverse Compton (IC), bremsstrahlung, and $\pi^0$-decay processes within leptonic and hadronic scenarios that are described below.
\subsection{Leptonic scenario}\label{sec:leptonic}
 
We first consider a leptonic scenario, i.e., the observed gamma-ray radiation at MeV--GeV energies is resulting from emission from relativistic electrons through inverse Compton (IC) and non-thermal bremsstrahlung processes. For simplicity, we consider a single population of electrons, which follows a power-law distribution as a function of energy with a high-energy cutoff (ECPL) at $E_{max}$. The distribution of target photon density used for the IC emission is described in Appendix \ref{sec:target_photon}. The ambient matter density ($n_0$) due to dense molecular clouds (MCs) around NGC 3603 was estimated to be about 1000 $\rm cm^{-3}$ by \citet{Fukui2014ApJ}. However \citet{Rui-zhi2017} calculated the average volume gas density to be $10 ~\rm cm^{-3} < n_0 < 60~ cm^{-3}$. Hence, for simplicity we consider ambient matter density of 35 $\rm cm^{-3}$, noting that larger or smaller values of the ambient matter density simply scale the contribution of the bremsstrahlung spectrum. 
Fig. \ref{fig:leptonic} shows both IC and bremsstrahlung spectra for the ECPL electron distribution, and it is evident that the bremsstrahlung spectrum can explain the observed SED at energies below 10 GeV. On the other hand, the IC emission for the target photons of Cosmic Microwave Background (CMB) and starlight can explain the observed SED in the energy range between 10 GeV to 100 GeV for the same population of electrons.

To check the contribution of the synchrotron spectrum for the electron distribution estimated above, we consider the X-ray spectrum above 3 keV where it follows a power-law spectral shape as discussed in Section \ref{sec:x-ray}. The association of non-thermal origin of the observed emission above 3 keV is not yet established. Hence, the total X-ray emission above 3 keV can be considered as flux upper-limit for the non-thermal emission. If its origin is associated with the electron distribution used for IC and bremsstrahlung spectra, then its contribution to the synchrotron spectrum is required to be estimated. Moreover, the fluxes from this emission should not overestimate the fluxes at X-ray energies. We find that synchrotron spectrum for value of magnetic field of about 10 $\mu G$ is neither able to explain the observed SED nor overestimates the fluxes at X-ray energies. Therefore, the magnetic field of this region should be lower than 10 $\mu G$.

\subsection{Hadronic scenario}\label{sec:hadronic}
\begin{figure}
\includegraphics[width=0.45\textwidth]{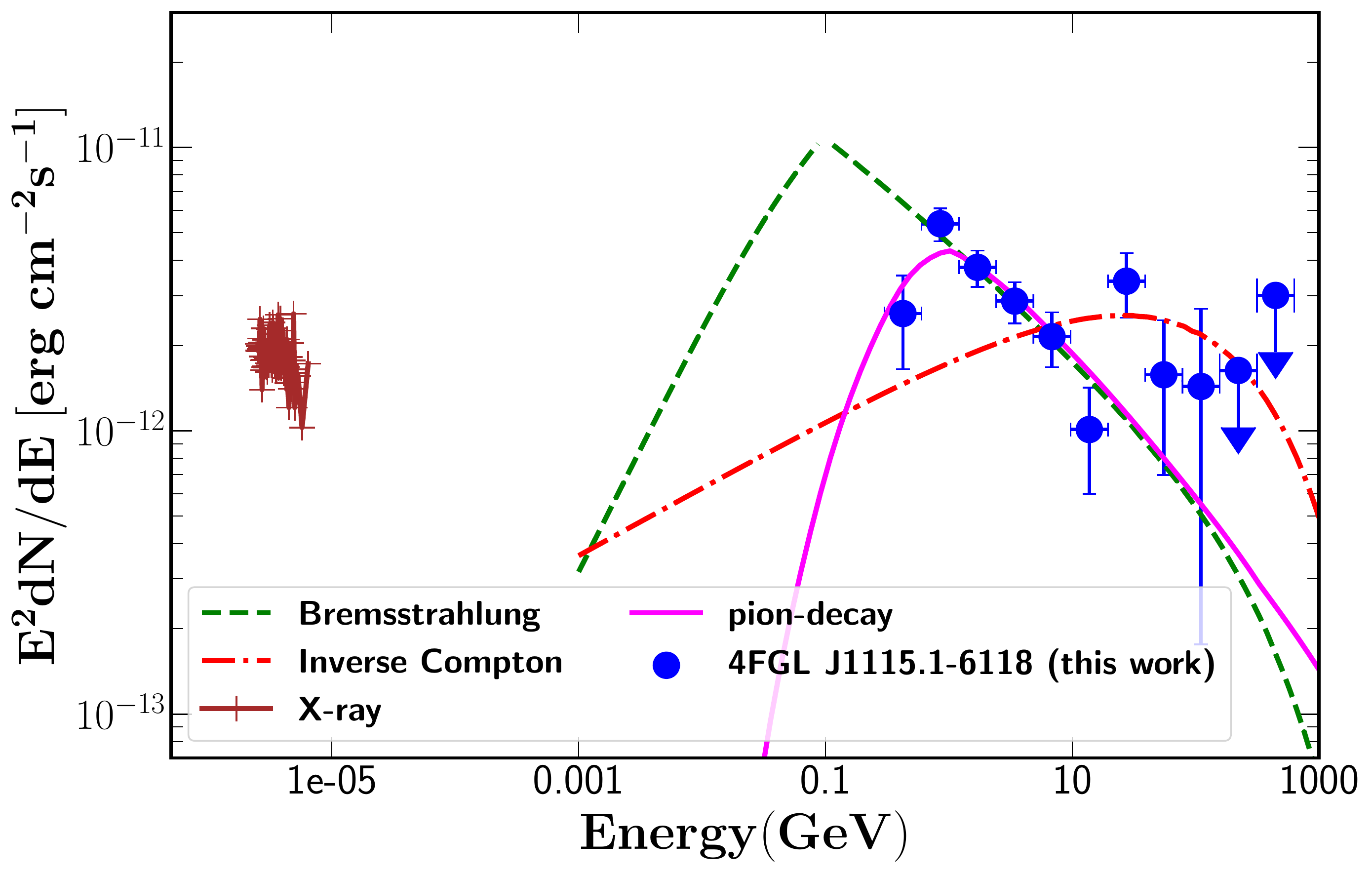}
\caption{The spectral energy distribution (SED) of \fhl\ for a leptonic model. A single zone electron distribution IC spectrum for the target photon distribution is shown (red dot-dashed line). The corresponding bremsstrahlung spectrum for an ambient matter density of 35 $\rm cm^{-3}$ is also shown (green dashed line). The gamma-ray spectrum (magenta solid line) resulting from the decay of $\pi^0$s is estimated for the ambient matter density of 35 $\rm cm^{-3}$. The parameters of the model are given in Table \ref{tab:fit_parameters}\label{fig:leptonic}. The error bars shown with the data points are statistical only. The X-ray SED data correspond to energies in the range of 2-10 keV.}
\end{figure}

To explore a hadronic origin of the observed SED, we estimate the gamma-ray flux resulting from the decay of neutral pions \citep[$\pi^0$s,][]{Kelner_2006} for a ECPL distribution of protons. This has similar spectral type as the one considered for the leptonic scenario. The ambient proton density is also considered to be $35~\rm {cm^{-3}}$. The SED for the hadronic scenario is shown in Fig. \ref{fig:leptonic}. The best-fit parameters of the model are shown in Table \ref{tab:fit_parameters}. Since, $\pi^0$-decay fluxes are proportional to the ambient matter density, a larger value of the density will reduce the total energy budget for the protons to explain the observed data. Fig. \ref{fig:leptonic} shows that the observed spectrum can be explained well with the hadronic scenario. The parameters of the models (Table \ref{tab:fit_parameters}) are not well constrained due to the large uncertainties in the SED at high energies ( $\gtrsim$ 20 GeV). 

Although it appears that the SED between 300 MeV to approximately 2 GeV is associated with the so-called pion-bump, a data analysis below 300 MeV would be required to confirm the presence of the break and reveal the signature of the ``pion-decay'' bump. This low-energy analysis in such a confused region is out of the scope of the paper. In addition, at energies above 10 GeV, a different spectral signature appears to be present. However, as discussed in Section \ref{sec:spectrum}, the spectral curvature is not significant. Nevertheless, we check the possibility of having more than one proton distribution within the emission region. We add another PL proton distribution to the ECPL proton distribution. We find that the best-fit results are not significantly better than a single ECPL model. The conclusion of this test is that the presence of an additional particle distribution can not be confirmed within the present scenario.

\begin{table*}
\caption{Parameters for physical models for a single zone particle distribution. The parameters are obtained  considering two different scenarios: leptonic and hadronic. }
\label{tab:fit_parameters}
\centering
\begin{tabular}{c|c|c}
\hline
\hline
Parameters &     Leptonic & Hadronic \\
           &              &          \\
\hline 
spectral index  ($\alpha$)                 &   2.5              & 2.3      \\
Low energy cutoff, $E_{min}$ (GeV)         &   $10^{-3}$         & 1.0        \\
High energy cutoff, $E_{cutoff}$ (GeV)     &   $1.0 \times 10^2 $               & 50 $\times 10^3$   \\ 
Ambient proton density, $n_0$ ($cm^{-3}$)  &   35                 & 35            \\
Total energy      ($10^{48}$ ergs)         &  4.6                & 5.5      \\
 \hline
 \hline
\end{tabular}
\end{table*}

\begin{figure}
\includegraphics[width=.45\textwidth]{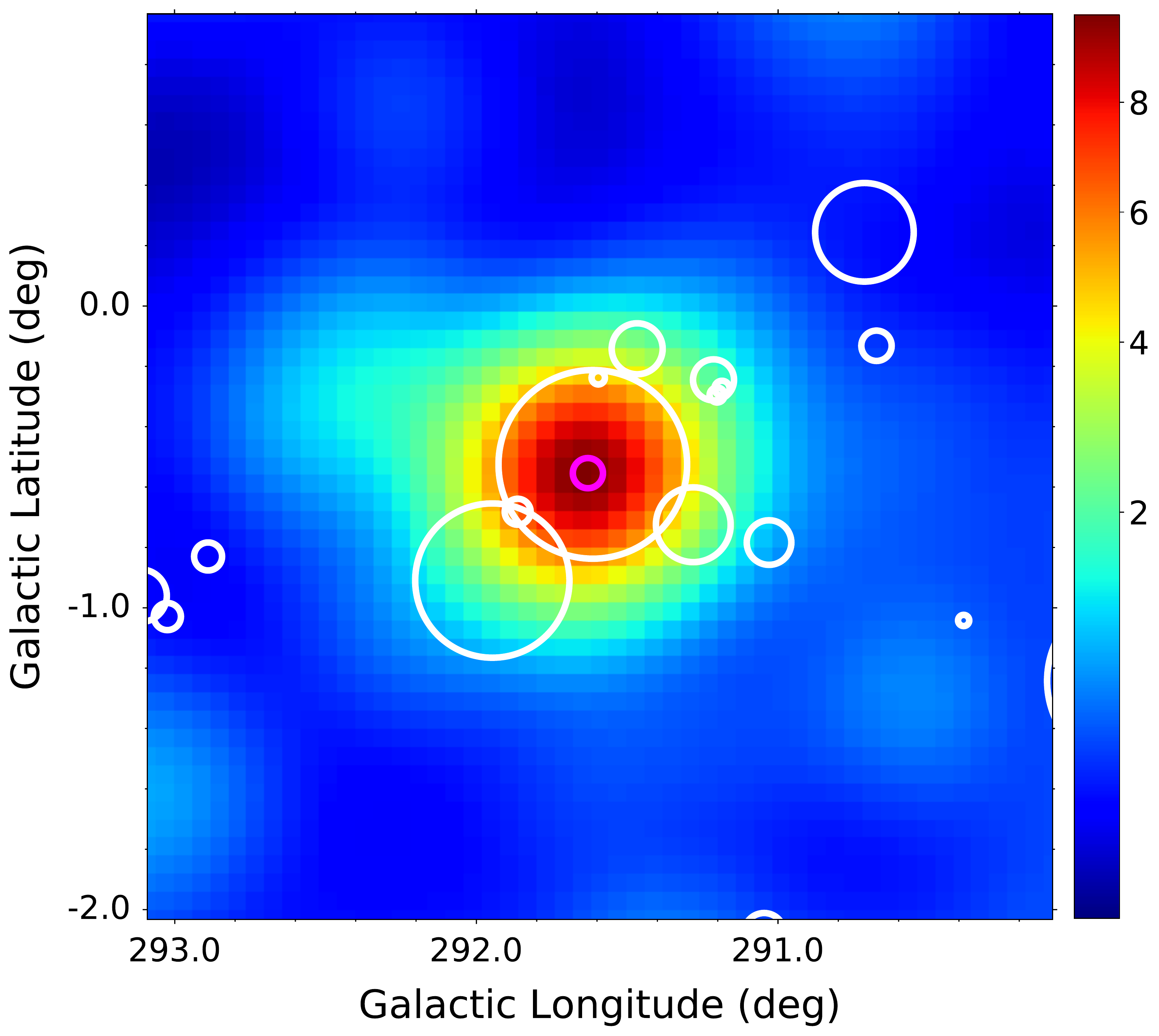}
\caption{The TS map of \fhl\ for \textit{Fermi}-LAT data for energy 10 GeV -- 1 TeV for \textit{model A} as discussed in Section \ref{sec:Gamma-ray data}. The H\textsc{ii} regions are shown with the white circles.  The GeV point source is well inside one of the H\textsc{ii} regions. The magenta circle indicates the 95\% positional uncertainty of \fhl.}\label{fig:hii_region}
\end{figure}

\begin{figure}
\includegraphics[width=.45\textwidth]{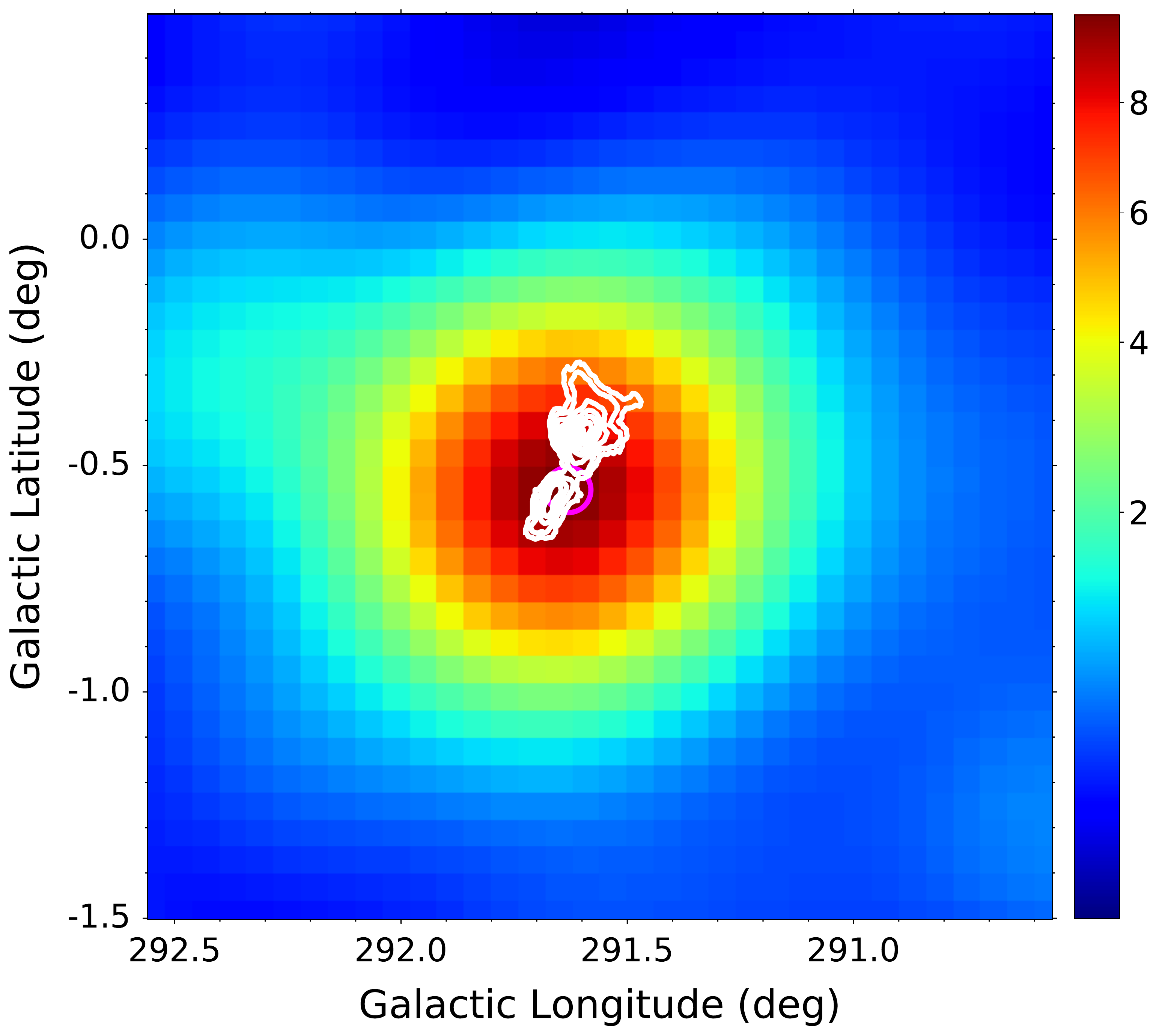}
\caption{Contours of  the molecular clouds \citep[white lines;][]{Fukui2014ApJ} are overlapped on the TS map of \fhl, which is similar to Fig. \ref{fig:hii_region} with reduced FOV. The magenta circle indicates the 95\% positional uncertainty radius of the point source \fhl.}\label{fig:MC_cloud}
\end{figure}

\section{Discussion} \label{sec:discussion}

\subsection{The nature of \fhl}
The \textit{Fermi}-LAT analysis of the data at MeV--GeV energies from the direction of NGC 3603 reveals that the observed gamma-ray emission is dominated by the pointlike source \fhl. In addition, a diffuse extended source covers a region of about $1^\circ$ that includes this pointlike source. The center of the extended emission is located at a distance of 0.5$^\circ$ from the center of the pointlike source. The diffuse emission region overlaps with some other pointlike sources as shown in Fig. \ref{fig:skymaps_sources}. Hence the diffuse emission might be associated with the possible contribution from all the sources present in this region.

The nature of \fhl\ remains uncertain.
\citet{moffat02} studied this region in detail using X-ray and radio data. A diffuse X-ray emission component is detected after removing the X-ray point sources from the ROI. The observed diffuse emission is about 20\% of the total observed X-ray flux from NGC 3603 and reaching out to a region with a radius of about 4 pc. However, the diffuse emission is thermal in nature and attributed to hot-star winds of a large number of merging or colliding stars or faint unresolved clusters \citep{moffat02}.

Radio observations at 3 cm and 6 cm did not show strong radio emission from this region except for two faint discrete radio sources. The observed radio emission is most likely non-thermal in origin and attributed to colliding winds. Moreover, the sizes of these radio sources were found to be $\sim 2''$. These sources are not associated with the brightest X-ray sources indicating that X-ray and radio emission mechanisms are not correlated. All these signatures are different from known young and old pulsar wind nebulae (PWNe). Therefore, it seems unlikely that \fhl\ is a PWN. Some of the exceptionally bright X-ray sources might be associated with colliding wind binaries. However, flux or spectral variability of these sources are not yet established, hence making them less likely candidates for gamma-ray binaries.

As reported in Section \ref{sec:x-ray}, we find 38 point-like bright X-ray sources within the \textit{Fermi}-LAT uncertainty region. The majority of the pointlike sources are associated with the young cluster. The bright X-ray sources analyzed in the ROI show spectral features consistent with what is expected for them to be O or B stars. 

We also look for sources that might be associated with extragalactic sources. It appears that the ROI contains a few X-ray emitting extragalactic sources (which is expected, given the large ROI), but none of them is bright enough or presents multiwavelength properties (e.g., presence of jets) that strongly suggest that the gamma-ray emission is of extragalactic origin. In our X-ray analysis we also find that leaving the column density N${\rm_H}$ free to vary does not significantly improve the spectral fit for all the sources fitted with a thermal model (i.e., OB stars). This evidence further supports the claim that all the sources, even those without a certain counterpart, may belong to the stellar cluster NGC 3603.

Recently the H.E.S.S. Collaboration has discovered very high energy gamma-ray emission from HESS J1119--614. This source is identified as the composite SNR G292.2--0.5 and associated with PWN G292.15--0.54 and the highly magnetized pulsar PSR J1119--6127 \citep{2018_HESS}.  We also look into the association of the observed emission from HESS J1119-614 with the \fhl\ source. However, we find that the centroids of this two objects are approximately $0.5^\circ$ away from each other. The 1$\sigma$ localization uncertainty radii of these sources also do not overlap with each other. The pulsar PSR J1119--6127 is associated with 4FGL 1119.1--6127,  which is shown in Fig. \ref{fig:skymaps_modelA}. In addition, the SED of HESS J1119-614 is fitted well with a power-law spectrum with spectral index 2.64 $\pm$ 0.12, which is steeper than that of \fhl\ that is 2.35 $\pm$ 0.03. Hence, we rule out the possible association of \fhl\ with the HESS source. There are a few more known pulsars present in this region, but they all are at a distance of 0.3 degrees or larger from the location of \fhl. In addition, no PWNe associated with these pulsars are known. Thus, a PWN association is not completely ruled out but it is a less likely candidate of \fhl.

The X-ray and radio data also confirm the absence of any shell-like morphology from the source. This also disfavors the possibility that \fhl\ is a supernova remnant. However, the total energy of electrons or protons of $\sim 10^{49}$ erg (see Table \ref{tab:fit_parameters}), which is $< 10\%$ of the energy budget of a supernova remnant, would make the hypothesis of a single unknown supernova remnant energetically viable. Conversely, the association of the observed emission with a pulsar can be excluded due to the presence of high energy photons extending more than 100 GeV. Pulsar emission drops off rapidly for energies above the spectral energy density peak around a few GeV \citep{MAGIC_Crab_2016A&A...585A.133A,Vela_2018A&A}. There is no spectral cutoff below 10 GeV in \fhl. Moreover, no significant variability is seen for any of the bright X-ray sources.
Therefore, we can speculate that the observed emission is associated with the SFR. The gamma-ray luminosity of the source is only $\sim$0.2\% of the total mechanical power of the winds from the SFR ($8.5 \times 10^{31}$ W, see Appendix \ref{app:winds} for details). Therefore energetically the hypothesis that the source is powered by the SFR is also acceptable. 

\subsection{Gamma rays from a star forming region}

The NGC 3603 region is observed in the infrared and found to have a clear bubble like structure, the so-called cavity. The radius of the cavity was estimated to be about approximately 1 pc \citep{Lacy1982}. One possible explanation for the presence of a bubble is  feedback from the SFR  on  MC and gas around the region by the winds from the stars, ionizing radiation and possibly SNR shocks. Hence the presence of a bubble-like structure is a good indication that the observed gamma rays could also be produced by this wind of accelerated particles. This is similar to what has been observed for the Cygnus cocoon in which the gamma-ray emission is confined in a bubble-like structure \citep{Ackermann2011Sci}.

Figure \ref{fig:hii_region} shows that the well-known H\textsc{ii} region \footnote{\textit{WISE} catalogue (V2.2) is used for the Galactic H\textsc{ii} regions (\url{http://astro.phys.wvu.edu/wise/})} covers the entire region of the \textit{Fermi}-LAT point source \fhl.  H\textsc{ii} regions are known to emit radiation through thermal bremsstrahlung (a.k.a free-free emission). They are expected to do so mainly due to the interaction of free electrons with ionized hydrogen present in the SFR. The intense radio emission suggests that massive OB association resides in the region and pushes the surrounding gas away to create a bubble.  The rich H\textsc{ii} region is the well-known characteristic of a SFR and it provides high potential ground for accelerating charged particles.  The H\textsc{ii} region ionized by the cluster is extended about 10 pc whereas the stellar winds cover a region of about 1 pc \citep{Claton90}.

\subsection{Molecular cloud environment}
\citet{Fukui2014ApJ} looked for the presence of potential MCs in this region of NGC 3603. They reported that  there are two of them present in this region and their centers are separated by about 10 pc at a distance of 7 kpc.  The authors claimed that these clouds most likely underwent some collisions that triggered the star formation. We overplot the contours of the MC with the \textit{Fermi}-LAT TS map in Fig. \ref{fig:MC_cloud}. It appears that the dominant contribution of gamma rays from the point source is coming not from the center of these MCs, but rather from the region in between these MCs. The gamma-ray emission region partially overlaps with the MC region \citep{Fukui2014ApJ}. However, the peak of the gamma-ray source cannot be resolved to the same resolution achieved by the CO map  due to lower angular resolution of \textit{Fermi}-LAT. Hence, a confirmed association with the peak of the gamma-ray emission and dense molecular clouds requires further detailed study such as energy dependent morphology, gamma-ray observations with higher angular resolution, etc. 

Figure \ref{fig:x-ray_map} shows that most of the bright X-ray sources are concentrated on the northern part of the 95\% localization uncertainty radius of the gamma-ray source. Therefore, from Fig. \ref{fig:x-ray_map} and Fig. \ref{fig:MC_cloud}  it can be understood that most of the bright X-ray point sources are well concentrated within the gap region of the MCs. In the SED fitting we consider relatively  low density of the ambient matter ($\sim 35~ {\rm cm^{-3}}$), which is supported by the low MC densities in this region. Also this low density of ambient matter indicates that the X-ray emission region is not obscured by the MC. Hence, the estimated N$\rm_H$ value is not getting contaminated due to MCs. The shift in the MC cloud density can possibly be associated with the transfer of kinetic energy of the winds to the MCs. The kinetic energy of this shift is estimated to be 10$^{47}$ ergs and can be supplied by the stellar winds having kinetic energy 10$^{51}$ erg for a 1 Myr timescale of the highest mass O stars \citep{Harayama2008ApJ}.  However, observed velocity distribution and cloud geometry also indicate that most of the cloud cannot be exposed to winds for the transfer of energy. 

The spectroscopic analysis of the stars in the very core of NGC 3603 showed that the stars provide more than 80\% of energy to ionize the gas in the surrounding nebula. However, the radius of the wind-driven nebula is reported to be much smaller despite its tremendous radiation energy \citep{Drissen1995}. Considering the huge available energy in the winds, they become a potential candidate for losing their energy through acceleration and radiation. We propose that the pointlike nature of the observed gamma-ray emission above 10 GeV might be associated with the wind-driven nebula and the energetic winds present in this volume is responsible for the observed photons above 10 GeV.

    
\section{Summary and Conclusions} \label{sec:conclusion}
We report the study of the young massive star cluster NGC 3603 at MeV-GeV energies using about 10 years of \textit{Fermi}-LAT data. The results are summarized below.

\begin{itemize}
    \item The results of our detailed analysis show that the observed gamma-ray emission from the \fhl\ source is not significantly extended.
    \item This region also contains the extended source \fges,  which plays a significant role in characterizing the morphological properties of the sources present in this region and require further studies.
    \item Observed X-ray emission is expected to be associated with the point source \fhl\ and Galactic in origin.
    
    \item The observed SED can be explained with both a leptonic and a hadronic model for density of ambient matter of $35~ {\rm cm^{-3}}$.
\end{itemize}

No firm evidence of association with any other classes of known gamma-ray emitters is found, therefore we speculate that 4FGL J1115.1-6118 is a case of gamma-ray emitting SFR.
Hence, it becomes a potential candidate for studying SFRs to understand the origin of cosmic rays using the next generation of gamma-ray telescopes such as the Cherenkov Telescope Array and the Major Atmospheric Cherenkov Experiment.

\section*{Acknowledgements}
L.S. acknowledges the research grant supported by the ERDF under the Spanish MINECO (FPA2015-68378-P and FPA2017-82729-C6-3-R).
A.D. thanks the support of the Ram\'on and Cajal program for the Spanish MINECO. S.M acknowledges support from the agreement ASI-INAF n.2017-14-H.O. We thank the anonymous referee for helpful comments.

The \textit{Fermi} LAT Collaboration acknowledges generous ongoing support
from a number of agencies and institutes that have supported both the
development and the operation of the LAT as well as scientific data analysis.
These include the National Aeronautics and Space Administration and the
Department of Energy in the United States, the Commissariat \`a l'Energie Atomique
and the Centre National de la Recherche Scientifique / Institut National de Physique
Nucl\'eaire et de Physique des Particules in France, the Agenzia Spaziale Italiana
and the Istituto Nazionale di Fisica Nucleare in Italy, the Ministry of Education,
Culture, Sports, Science and Technology (MEXT), High Energy Accelerator Research
Organization (KEK) and Japan Aerospace Exploration Agency (JAXA) in Japan, and
the K.~A.~Wallenberg Foundation, the Swedish Research Council and the
Swedish National Space Board in Sweden. Additional support for science analysis during the operations phase is gratefully acknowledged from the Istituto Nazionale di Astrofisica in Italy and the Centre National d'Etudes Spatiales in France. This work performed in part under DOE Contract DE- AC02-76SF00515.

\software{fermipy \citep[v0.17.4;][]{Wood_2017ICRC...35..824W}, Fermitools (Fermi Science Support development Team 2019), CIAO \citep[v4.9;][]{fruscione06}, XSPEC \citep{Arnaud_1996ASPC..101...17A}}

\appendix

\section{Results on X-ray data analysis}\label{sec:x-ray_table}
The \texttt{wavdetect} tool of CIAO is used to identity point-like and moderately extended X-ray sources. We perform our analysis using four different wavelet scales, i.e., 1, 2, 4 and 8 pixels. We exclude from our analysis the central part of the cluster, marked in Figure \ref{fig:x-ray_map} with an ellipse (green), because this part of the observation is significantly affected by pileup\footnote{\url{http://cxc.harvard.edu/ciao/why/pileup_intro.html}} (pileup fraction $f_p>$10\,\% in $>$30\,\% of the elliptic region), thus making any spectral measurement unreliable. Overall, we find 300 sources, covering a wide range of measured net counts in the 0.5--7\,keV band: the faintest source has 4.7 net counts and the brightest 4130.
There are 38 sources with net counts more than 70 and they are therefore bright enough to perform a basic spectral fitting \citep[see, e.g.,][]{marchesi16}. Given the relatively low counts  of most of the objects in our sample, we use the W statistic, i.e., the version of the Cash statistic  used when a background spectrum is available, binning each spectrum with at least 3 counts per bin \citep{cash79}. In all fits, we fix the absorption value to the Galactic one (N$_{\rm H,Gal}$=1.16 $\times10^{22}\,\rm cm^{-2}$), since we find that the fits are not significantly improved by leaving the parameter free to vary.
We do not fit the data of the brightest source in our sample (the OB star NGC 3603 47) because we find it to be significantly affected by pile-up ($f_p>$5\,\%).

Before performing the X-ray spectral analysis, we identify the counterparts of our bright X-ray sources, with a cross-match, allowing a maximum distance of 2$^{\prime\prime}$ between the X-ray and the optical position. We find a counterpart for 30 out of 38 objects: 25 are stars, one is a blend of two stars, and the remaining four are infrared (IR) sources.

 In Table \ref{tab:results_mekal} we report the best-fit results for the 30 sources best-fitted with a thermal \texttt{mekal} model \citep{mewe85,mewe86,liedahl95}, which is commonly adopted to fit the X-ray spectra of young, bright stars. All but one of the sources are fitted with a single-temperature model: the average temperature is $\langle kT \rangle$=1.72\,keV, with a standard deviation $\sigma_{\rm kT}$=0.98\,keV. In NGC 3603 56, instead, we find that the fit is significantly improved by the addition of a second thermal component. In most of the cases, we fix the metallicity value $Z$ to Solar: however, in seven sources we measure a significant improvement in the fit statistic when leaving $Z$ free to vary.
 
Table \ref{tab:results_PL} reports the best-fit results for the other seven objects best-fitted with a power-law model: four of these sources are unassociated, two are IR sources, and one is a blend of two stars; for this last source, the power-law best fit model should be interpreted as a phenomenological fit to two blended thermal components. The unassociated objects are likely not stars and can be extragalactic sources, possibly active galactic nuclei. Particularly, in two objects (sources 7 and 34) the fit is significantly improved by the addition of an absorption component (N$_{\rm H,l.o.s.}$; \texttt{pha} in XSPEC), possibly caused by the obscuring torus located nearby the accreting supermassive black hole powering the active galactic nuclei. However, no multiwavelength properties are observed from them. Moreover, the lack of redshift information for these two objects does not allow us to properly constrain $\Gamma$ and N$_{\rm H,l.o.s.}$.

The X-ray spectrum is also estimated from the region of \fhl. Given the relatively large size of the \fhl\ region with respect to the Chandra spatial resolution, this X-ray spectrum is the combination of multiple unresolved sources and the model we used is thus purely phenomenological. 
The spectrum is best fitted with a model \texttt{pha*(pow+mekal+gauss)}, where \texttt{pha} is the Galactic NH (1.16 $\times10^{22}\,\rm cm^{-2}$), \texttt{pow} is a power-law with spectral index 2.3, \texttt{mekal} is a thermal component with kT=2.6 and Solar metallicity, and \texttt{gauss} is a Gaussian at E=2.42 keV.

\begingroup
\renewcommand*{\arraystretch}{1.15}
\begin{table*}
\centering
\begin{tabular}{ccccccccc}
\hline
\hline
  ID & RA       & DEC      & cts$_{\rm 0.5-7}$	& Counterpart		& kT${_1}$		& $Z_{1}$				& kT$_{2}$		& CStat/d.o.f. \\
    &   deg     &   deg      &                    &                   & keV           & Z$_{\odot}$   &   
    keV         & \\
\hline
  1  & 168.7890 & -61.2673 & 4128.39   	& NGC 3603 47		& --                      & --                                	  	& --			& -- \\
  2  & 168.7785 & -61.2600 & 1175.63   	& [HEM2008] 51 		& 1.02$_{-0.07}^{+0.07}$  & 0.28$_{-0.10}^{+0.15}$	& --			& 169.6/176 \\
  3  & 168.7920 & -61.2606 & 646.39    	& NGC 3603 22		& 0.93$_{-0.09}^{+0.09}$  & 0.29$_{-0.14}^{+0.31}$	& --			& 119.3/109 \\
  4  & 168.7863 & -61.2667 & 634.89    	& NGC 3603 18		& 1.29$_{-0.07}^{+0.07}$  & 1.00$^{f}$			& --			& 118.8/119 \\
  5  & 168.7816 & -61.2819 & 453.56    	& [SB2004] 1267		& 1.30$_{-0.15}^{+0.18}$  & 0.07$_{-0.07}^{+0.16}$	& --			& 120.9/103 \\
  6  & 168.7841 & -61.2632 & 386.65    	& NGC 3603 57		& 0.72$_{-0.12}^{+0.10}$  & 0.35$_{-0.21}^{+0.65}$	& --			& 68.3/76 \\
  8  & 168.7810 & -61.2635 & 279.55    	& [HEM2008] 171		& 4.41$_{-0.96}^{+1.59}$  & 1.00$^{f}$			& --			& 90.4/90 \\
  9  & 168.7759 & -61.2602 & 255.02    	& [HEM2008] 19 		& 0.92$_{-0.14}^{+0.08}$  & 1.00$^{f}$			& --			& 70.1/65 \\
  10 & 168.7769 & -61.2602 & 207.35     & [HEM2008] 76		& 2.26$_{-0.32}^{+0.64}$  & 1.00$^{f}$			& --			& 61.8/58 \\
  11 & 168.7802 & -61.2597 & 201.43    	& [HEM2008] 9		& 0.81$_{-0.07}^{+0.09}$  & 1.00$^{f}$			& --			& 87.6/66 \\
  12 & 168.7813 & -61.2629 & 190.33    	& NGC 3603 56		& 0.98$_{-0.58}^{+0.36}$  & 1.00$^{f}$			& 3.91$_{-1.52}^{+6.42}$	& 66.8/56 \\
  13 & 168.7824 & -61.2578 & 189.49    	& NGC 3603 64		& 1.02$_{-0.09}^{+0.09}$  & 1.00$^{f}$			& --			& 75.4/50 \\
  14 & 168.7776 & -61.2613 & 180.27    	& [HEM2008] 136		& 1.65$_{-0.20}^{+0.35}$  & 0.19$_{-0.19}^{+0.40}$	& --			& 91.0/69 \\
  15 & 168.7794 & -61.2582 & 174.10     & [HEM2008] 249		& 3.29$_{-0.71}^{+1.15}$  & 1.00$^{f}$			& --			& 62.7/54 \\
  16 & 168.7822 & -61.2598 & 173.77    	& [HEM2008] 103		& 1.02$_{-0.09}^{+0.09}$  & 1.00$^{f}$			& --			& 75.4/50 \\
  17 & 168.7858 & -61.2604 & 149.42    	& [HEM2008] 105		& 1.76$_{-0.38}^{+0.68}$  & 0.10$_{-0.10}^{+1.07}$	& --			& 30.7/38 \\
  18 & 168.7673 & -61.2606 & 144.35    	& NGC 3603 54		& 2.82$_{-0.58}^{+1.14}$  & 1.00$^{f}$			& --			& 40.9/42 \\	
  19 & 168.7779 & -61.2590 & 136.06    	& [HEM2008] 46		& 1.62$_{-0.37}^{+0.49}$  & 0.30$_{-0.28}^{+0.94}$	& --			& 53.6/41 \\
  20 & 168.7779 & -61.2760 & 135.33    	& UCAC4 144-073112	& 0.81$_{-0.17}^{+0.16}$  & 1.00$^{f}$			& --			& 43.7/34 \\
  21 & 168.7972 & -61.2655 & 135.25    	& NGC 3603 19		& 0.60$_{-0.12}^{+0.13}$  & 1.00$^{f}$			& --			& 32.9/30 \\
  22 & 168.7722 & -61.2586 & 132.52    	& [SB2004] 10018	& 2.53$_{-0.56}^{+0.92}$  & 1.00$^{f}$			& --			& 27.5/36 \\
  25 & 168.7978 & -61.2677 & 98.58     `& None			& 1.56$_{-0.28}^{+0.32}$  & 1.00$^{f}$			& --			& 14.7/29 \\
  26 & 168.7881 & -61.2593 & 93.61     	& NGC 3603 49		& 1.59$_{-0.33}^{+0.31}$  & 1.00$^{f}$			& --			& 48.2/38 \\	
  28 & 168.8589 & -61.3017 & 88.33     	& [SB2004] 57862	& 0.51$_{-0.15}^{+0.14}$  & 1.00$^{f}$			& --			& 46.1/28 \\
  29 & 168.7628 & -61.2657 & 88.14     	& [SB2004] 20701	& 3.24$_{-1.25}^{+4.68}$  & 1.00$^{f}$			& --			& 22.9/21 \\
  30 & 168.7787 & -61.2626 & 88.0     	& NGC 3603 63		& 0.51$_{-0.10}^{+0.12}$  & 1.00$^{f}$			& --			& 37.4/30 \\
  31 & 168.7667 & -61.2635 & 87.1     	& [NS2003] 7		& 1.31$_{-0.31}^{+0.60}$  & 1.00$^{f}$			& --			& 20.7/21 \\
  32 & 168.7750 & -61.2634 & 79.2     	& [NS2003] 8		& 2.62$_{-0.60}^{+1.07}$  & 1.00$^{f}$			& --			& 24.3/30 \\
  33 & 168.8109 & -61.2652 & 76.2     	& None			& 2.16$_{-0.50}^{+1.00}$  & 1.00$^{f}$			& --			& 14.3/20 \\
  35 & 168.7908 & -61.2691 & 74.9     	& None			& 3.19$_{-1.03}^{+2.54}$  & 1.00$^{f}$			& --			& 25.3/20 \\
  37 & 168.7933 & -61.2738 & 70.9     	& None			& 2.61$_{-0.68}^{+1.38}$  & 1.00$^{f}$			& --			& 15.4/21 \\
\hline
\hline
\end{tabular}
\caption{Summary of the best-fit results for the 31 out of 38 x-ray sources with more than 70 net counts (cts$_{\rm 0.5-7}$) in the 0.5-7 keV band, best fitted with a thermal \texttt{mekal} model. We do not report the results relative to source 1 because its spectrum is significantly affected by pile-up. kT${_1}$ and kT${_2}$ are the temperature of the first and second thermal components, respectively, whereas Z$_{1}$ is the metallicity of the first thermal component. Parameters flagged with $^f$ were frozen in the fit.}\label{tab:results_mekal}
\end{table*}
\endgroup

\begingroup
\renewcommand*{\arraystretch}{1.15}
\begin{table*}
\centering
\centering
\begin{tabular}{ccccccc}
\hline
\hline
  ID & RA       & DEC      & cts$_{\rm 0.5-7}$	& Counterpart		& $\Gamma$		& CStat/d.o.f. \\
    &   deg     &   deg  \\
\hline
  7  & 168.7517 & -61.2765 & 305.5    	& None					& Obscured Power Law		& 	      \\
  23 & 168.7815 & -61.2558 & 121.6    	& [NS2003] 6H				& 1.72$_{-0.36}^{+0.36}$	& 34.4/35 \\		
  24 & 168.7880 & -61.2692 & 107.8     & None					& 2.06$_{-0.37}^{+0.37}$	& 28.3/34 \\
  27 & 168.7832 & -61.2595 & 89.6     	& [HEM2008] 148,[HEM2008] 179(blend)	& 2.38$_{-0.35}^{+0.35}$	& 40.9/38 \\
  34 & 168.8014 & -61.2931 & 76.0     	& None					& Obscured Power Law		& 	  \\
  36 & 168.7989 & -61.2780 & 73.0     	& [TBP2001] J111511.8-611641		& 1.71$_{-0.55}^{+0.55}$	& 17.4/16 \\
  38 & 168.8240 & -61.2807 & 70.0     	& None					& 2.02$_{-0.50}^{+0.50}$	& 26.2/17 \\
\hline
\hline
\end{tabular}
\caption{Summary of the best-fit results for the 7 out of 38 x-ray sources with more than 70 net counts in the 0.5-7 keV band best fitted with a non-thermal power-law model. cts$_{\rm 0.5-7}$ are the net counts in the 0.5--7\,keV band and $\Gamma$ is the power-law photon index. Parameters flagged with $^f$ were frozen in the fit. Source 7 and 34 are best-fitted by and obscured power-law model, but the correct $\Gamma$ and $N_{\rm H,l.o.s.}$ values cannot be determined without knowing the source redshift.}\label{tab:results_PL}
\end{table*}
\endgroup

\section{Inverse Compton target radiation field}\label{sec:target_photon} \label{sec:target_field}
\begin{figure*}
\includegraphics[width=.45\textwidth]{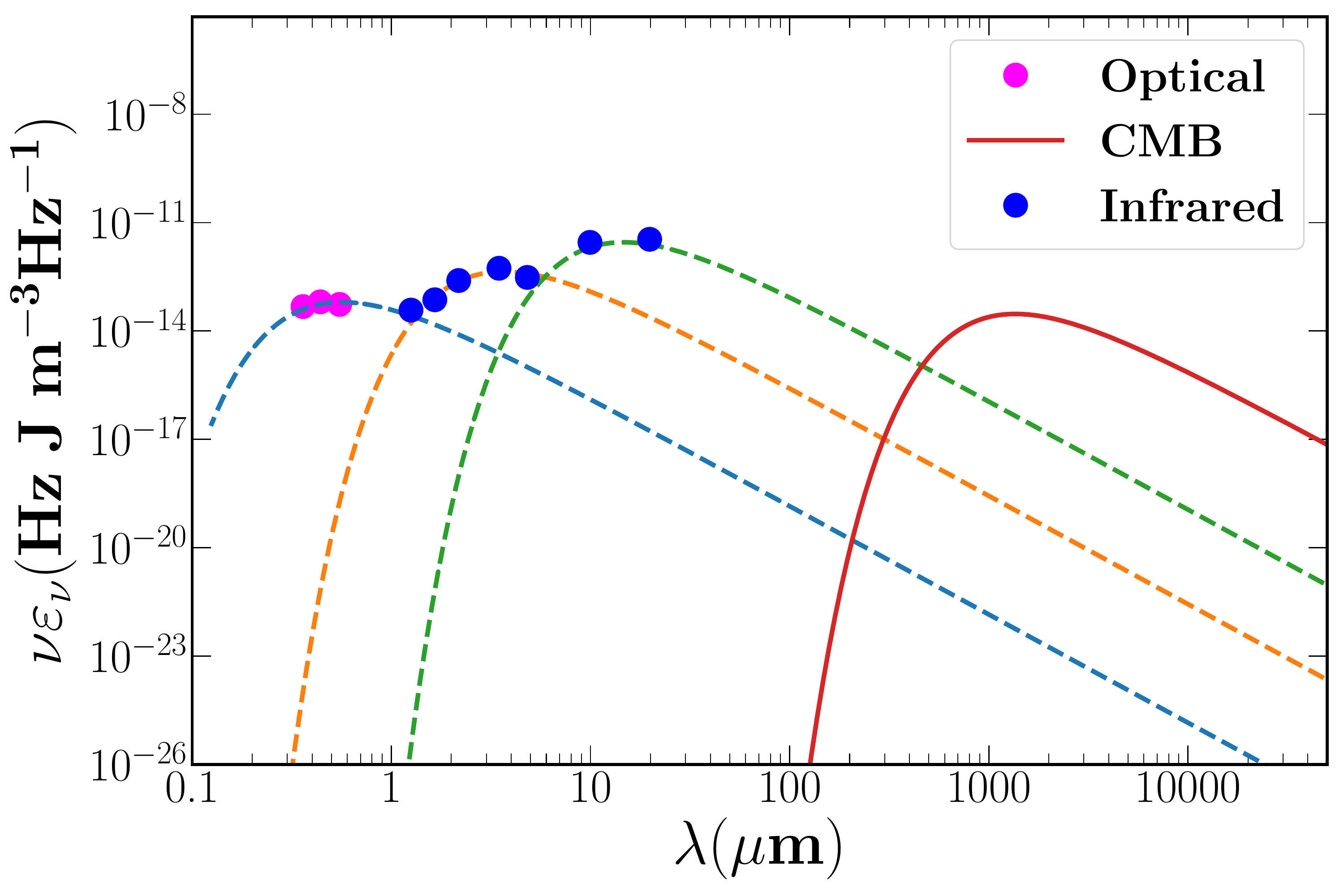}
\caption{The target photon energy density of the soft radiation, which are modelled using black body spectra. It includes optical, near infrared, mid infrared radiation and CMB photons. \label{fig:ISRF}}
\end{figure*}

We calculate the target photon density for the IC emission process using published results from the direction NGC 3603. \citet{Nurmberger_2002A&A} studied the NGC 3603 region, focusing on the brightest source IRS 9 in the near infrared and mid infrared region using the Las Campanas 2.5 m telescope.  The center of the bright IRS 9 star is located 1\arcmin.2 south of the center of the OB cluster. The spectral energy distribution at these energy bands for different parts of the IRS 9 were modelled with a combination of black body spectra. We consider spectral data points for the brightest part of the region, which are modelled with a black body spectrum at 250 K and 1000 K \citep{Nurmberger_2002A&A}. The contribution to the radiation field by optical observations made by \citet{OpticalNGC36031965} is also considered. We model the observed optical photon flux with a black body spectrum for a temperature of 6500 K. The target photon density at the source is estimated considering the source distance 7 kpc and angular extension of these optical and infrared observations. The photon densities are measured to be 24.21 eV $\rm cm^{-3}$, 3.70 eV $\rm cm^{-3}$, 0.53 eV $\rm cm^{-3}$ with the associated temperatures 250 K, 1000 K and 6500 K, respectively. In addition to these two different distributions of target photons, we consider cosmic microwave background (CMB) radiation for the IC contribution to the observed spectrum. Figure \ref{fig:ISRF} presents the spectrum of the average target photon energy density used to calculate the IC spectrum. 

\section{Stellar wind properties in NGC~3603}\label{app:winds}

\subsection{Stellar populations in the cluster}\label{stellarpar}

A stellar population model is built for NGC 3603 following \citet{Ackermann2011Sci}. We consider that star counts distributed in the cluster as $dn/d M_* \propto M_*^{-\alpha}$ with $\alpha=-1.73$, normalized so that we have a total of 28 stars in the O5 to O3 classes \citep{Harayama2008ApJ}. We distribute star counts in four bins, namely, B8 to B5, B5 to B0, O9 to O5 and O5 to O3. The sample is limited to stars heavier than B8 due to the validity range for the reference mass-loss rate model adopted in Section~\ref{starwinds}; in Section~\ref{clusters} we will see that the contribution to the mechanical energy injection rate and global mass loss from lighter stars is negligible.

In each bin we can take a representative value for the stellar mass $M_*$, luminosity $L_*$, effective temperature $T_*$ and surface radius $R_*$. We adopt typical values for supergiants (class~I) O stars \citep{Martins2005A&A} and for main-sequence (class~V) B stars \citep{Cox2000asqu.book}. The representative average value is derived assuming a power-law distribution as a function of mass for each spectral interval. All values are reported in Table~\ref{cygob2}.

\subsection{Individual stellar wind properties}\label{starwinds}

Based on the representative stellar parameters introduced in Section~\ref{stellarpar} we can assign to each spectral type a representative wind terminal velocity $v_{\infty \; *}$ and mass loss $\dot{M}_*$ by following the prescriptions of a semi-empirical model \citep{Vink2000A&A}.

First, according to the stellar parameters we assign each spectral type to one side of the \emph{bi-stability jump} due to the drastic change in the wind ionization occurring at $\sim 25000$~K: for our representative values O stars are situated on the hot side the jump and B stars on
the cold one. We then calculate $v_{\infty \; *}$ and $\dot{M}_*$. The results are also reported in Table~\ref{cygob2}.

\begin{table}
\begin{center}
\caption{Star counts and representative stellar parameters for the NGC~3603 cluster.}\label{cygob2}
\begin{tabular}{lcccc}
\hline
					& B8--B5 	& B5--B0 	& O9--O5 	& O5--O3 \\
\hline
counts					& $142$		& $204$		& $78$	& $28$ \\
$M_*$ ($M_\odot$)			& $4.7$	& $9.9$	& $24.5$	& $47.1$ \\
$L_*$ ($\log_{10}{L_*/L_\odot}$)	& $2.7$	& $4.1$	& $5.7$	& $6.0$ \\
$R_*$ ($R_\odot$)			& $3.4$	& $5.2$	& $22.4$	& $19.0$ \\
$T_*$ (K)				& $13100$	& $20800$	& $32700$	& $40400$ \\
$v_{\infty \; *}$ (km/s)		& $945$		& $1105$	& $1680$	& $2526$ \\
$\dot{M}_*$ ($M_\odot$/year)		& $2.3\times 10^{-11}$	& $1.7\times 10^{-8}$	&
$4.4\times 10^{-6}$	& $9.8\times 10^{-6}$\\
\hline
\end{tabular}
\end{center}
\end{table}

\subsection{Cluster properties}\label{clusters}
The total mass injection rate for the cluster is
$$
\dot{M}_{SC}=\sum_* \, \dot{M}_*
$$
The total mechanical energy injection rate is
$$
\dot{E}_{SC}=\sum_* \, \frac{1}{2} \dot{M}_* v_{\infty \; *}^2
$$
From the values in Table~\ref{cygob2} one can see that the contribution to both
the mass and the mechanical energy injection rates from stars lighter than B5 is negligible.

If we assume that the central overpressure drives mass away from the cluster in the form of a
coherent gaseous outflow, i.e. a stellar cluster wind, and we neglect radiative energy
losses, once a steady-state is reached the cluster wind terminal velocity $V_{\infty\, SC}$
satisfies the equation
$$
\dot{E}_{SC}=\frac{1}{2} \dot{M}_{SC} V_{\infty\, SC}^2
$$
The (more complex) case of non-negligible radiative losses in the wind is treated, e.g., in
\citet{Silich2011ApJ}.

Cluster wind parameters obtained by applying the formulas above to the
stellar population parameters  given in Table~\ref{cygob2} are presented in
Table~\ref{clustertable}.

\begin{table}
\begin{center}
\caption{Total mass injection rate, mechanical energy injection rate and
terminal cluster wind velocity for NGC~3603.}\label{clustertable}
\begin{tabular}{lc}
\hline
$\dot{M}_{SC}$ ($M_\odot$/year)		& $6\times 10^{-4}$	 \\
$V_{\infty\, SC}$ (km/s)		& $2100$		 \\
$\dot{E}_{SC}$ (W)			& $8.5\times 10^{31}$	 \\
\hline
\end{tabular}
\end{center}
\end{table}

\bibliography{3fhl.bib}

\end{document}